\documentclass[twocolumn,reprint,amsmath,amssymb,aps,floatfix,superscriptaddress]{revtex4-2}
\usepackage{float}
\usepackage{graphicx}
\usepackage{amssymb}
\usepackage{mathtools}
\usepackage{dcolumn}
\usepackage{amsmath}
\usepackage{multirow}
\usepackage{makecell}
\usepackage{hhline}
\usepackage[justification=Justified]{caption}
\usepackage{subcaption}
\usepackage{xcolor}
\usepackage{gensymb}
\usepackage[breaklinks]{hyperref}
\hypersetup{colorlinks,urlcolor=blue,citecolor=blue,linkcolor=blue,filecolor=blue}
\usepackage{anysize}
\usepackage[left=1.5cm,right=1cm,top=1cm,bottom=1cm,includehead,includefoot]{geometry}
\usepackage[skip=0pt plus0pt, indent=10pt]{parskip}
\usepackage[bottom]{footmisc}
\newcommand{\pref}[1]{(\ref{#1})}
\newcommand{\sref}[2]{\ref{#1}(\subref{#2})}
\usepackage[normalem]{ulem}
\newcommand{\mfrac}[2]{\scalebox{0.8}{$\dfrac{#1}{#2}$}}

\begin{document}
\title{Universal features of epidemic and vaccine models}
\author{Sourav Chowdhury}
\email{chowdhury95sourav@gmail.com}
\affiliation{Postgraduate and Research Department of Physics, St. Xavier's College (Autonomous) \\ 30 Mother Teresa Sarani, Kolkata-700016, West Bengal, India}

\author{Indrani Bose}
\email{indrani@jcbose.ac.in}
\affiliation{Department of Physics, Bose Institute, Kolkata, India}

\author{Suparna Roychowdhury}
\email{suparna@sxccal.edu}
\affiliation{Postgraduate and Research Department of Physics, St. Xavier's College (Autonomous) \\ 30 Mother Teresa Sarani, Kolkata-700016, West Bengal, India}

\author{Indranath Chaudhuri}
\email{indranath@sxccal.edu}
\affiliation{Postgraduate and Research Department of Physics, St. Xavier's College (Autonomous) \\ 30 Mother Teresa Sarani, Kolkata-700016, West Bengal, India}
\begin{abstract}
    In this paper, we study a stochastic susceptible-infected-susceptible (SIS) epidemic model that includes an additional immigration process. In the presence of multiplicative noise, generated by environmental perturbations, the model exhibits noise-induced transitions. The bifurcation diagram has two distinct regions of unimodality and bimodality in which the steady-state probability distribution has one and two peaks, respectively. Apart from first-order transitions between the two regimes, a critical-point transition occurs at a cusp point with the transition belonging to the mean-field Ising universality class. The epidemic model shares these features with the well-known Horsthemke-Lefever model of population genetics. The effect of vaccination on the spread/containment of the epidemic in a stochastic setting is also studied. We further propose a general vaccine-hesitancy model, along the lines of Kirman’s ant model, with the steady-state distribution of the fraction of the vaccine-willing population given by the Beta distribution. The distribution is shown to give a good fit to the COVID-19 data on vaccine hesitancy and vaccination. We derive the steady-state probability distribution of the basic reproduction number, a key parameter in epidemiology, based on a beta-distributed fraction of the vaccinated population. Our study highlights the universal features that epidemic and vaccine models share with other dynamical models. 
\end{abstract}
\maketitle

\section{Introduction}\label{Sec:intro}

	The spread of an epidemic brought about by an infectious disease has close parallels in disciplines as diverse as biological, social, and information sciences \cite{vespignani2012,goffman1964,llyod2001}. Examples include the propagation of viruses in a population of cells, cultural norms in a society of individuals, and information in a communication network. Though the modeling context in each case is different, one can identify certain universal features associated with the spreading phenomenon. The simplest epidemic model, based on contagion dynamics, is the susceptible-infected-susceptible (SIS) model \cite{bailey1975,pastorsatorras2015}. In this model, a population of agents/individuals is divided into two categories: susceptible and infected. In the course of time, infected individuals infect susceptible individuals through contact, with the newly infected individuals participating in the further spreading of the infection. In the other key process of the model, an infected individual reverts back to the susceptible state without acquiring immunity from the infection. The recovered individual may thus be reinfected at a future instant of time. In the SIS model, as defined above, a total eradication of the infection is possible. This is, however, no longer true on the inclusion of an immigration process in the SIS model \cite{oregan2013}. In this paper, we show that the dynamics of the SIS model with immigration have close similarities with the extensively studied Horsthemke-Lefever (HL) model of population genetics \cite{horsthemke1984,arnold1978,bose2022}, a connection not established earlier. In the presence of multiplicative noise, the HL model exhibits noise-induced transitions in the nonequilibrium.
    
    We study the stochastic SIS model with immigration to explore the effect of random fluctuations (noise) on the epidemic dynamics. The fluctuations, an outcome of stochastic processes, have both intrinsic and extrinsic origins. In the limit of large system (population) size, the internal fluctuations may be ignored as the noise intensity is inversely proportional to the system size. The external noise, arising from environmental stochasticity may, however, have a nontrivial effect on the dynamics. The effect of such stochasticity is to generate random fluctuations in the model parameters around their mean values. In the case of our stochastic SIS model, we introduce multiplicative noise (noise dependent on state variables) in the infection transmission parameter and obtain analytically the steady-state probability distribution. Two special cases of this distribution correspond to the HL model \cite{horsthemke1984,arnold1978,bose2022} and the SIS model without immigration \cite{mamis2023}. The bifurcation diagram of our stochastic SIS model shares universal features with that of the HL model, namely, the existence of a noise-induced transition and the presence of a cusp point at which a critical-point transition, belonging to the mean-field Ising universality class, occurs. Recent studies have explored the connection between bifurcation and mean-field Ising criticality in a number of stochastic models including models of biochemical positive feedback and a model describing the evolution of cancer \cite{erez2019,bose2019,bose2022criticality}.

    We point out at this stage that our study is carried out at a mean-field level and is limited to a single-scale, homogeneous population in which infections spread through pairwise contacts. Studies based on multiscale and spatially inhomogeneous models capture more accurately the complexity of the dynamics of disease transmission \cite{xian2025,sun2025,feld2022,fibich2024}. In multiscale modeling, influences operating at different scales, from local to global, are taken into account for an in-depth description of the global spread of an infectious disease \cite{xian2025}. Spatial inhomogeneity is taken into account by defining disease models on networks \cite{pastorsatorras2015} in which the contact structure for infection transmission between individuals is nonuniform, capturing a realistic situation. The highly connected individuals in a network spread the infection more effectively than an average individual whose contact structure is limited to a much smaller number of individuals. The inclusion of contact heterogeneity allows for a more accurate modeling of disease transmission. Reaction-diffusion models of disease spread integrate both spatial heterogeneity and individual movement within a local/global population \cite{sun2025}. Disease modeling, based on large-deviation approaches, opens up the possibility of probing the structure of the disease dynamics in terms of possible dynamical regimes including regimes marked by rare events, which are not accessible by standard simulation techniques \cite{feld2022}. Disease modeling on hypernetworks extends infection transmission beyond simple pairwise contacts \cite{fibich2024}.
    
    To further investigate the effect of stochasticity on the evolution of the epidemic, specifically, from the perspective of real-life situations, we include a vaccination parameter, providing a measure of the vaccinated fraction of the population, in the stochastic SIS model without immigration. The parameter modifies the average infection transmission rate through a reduced possibility of infection \cite{oregan2013}. From our analysis of the bifurcation diagram, we find that a combination of vaccination and stochasticity offers a potent strategy against infection. A vaccine-related issue of significant concern is the vaccine-hesitancy problem, i.e., the willingness/unwillingness of individuals in a population to be vaccinated \cite{franceschi2023}. We propose a model of vaccine hesitancy based on Kirman’s ant colony model \cite{kirman1993}, a stochastic model that illustrates the herding mechanism operating in a group of ants foraging for food from two identical food sources in the neighborhood. The model, based on binary decision-making processes, is of general applicability, e.g., in entomology, in financial markets displaying herding and contagion-related phenomena, and in the analysis of bipolar electoral voting patterns \cite{kirman1993,alfarano2005,kononovicius2018}. In our vaccine-hesitancy (denoted by VH in the rest of the paper) model along the lines of Kirman’s ant model, we show that the steady-state distribution of the vaccine-willing fraction of the population is described by the Beta distribution. The reliability of the model is tested by fitting the Beta distribution to available COVID-19 data on vaccine hesitancy \cite{lazarus2023}. We further derive the steady-state distribution of the basic reproduction number $R_{0}$, a key parameter in epidemiology, based on a Beta-distributed fraction of the vaccinated population. 

    The paper is organized as follows: Sec.~\ref{Sec:det_dyn} provides a brief description of the deterministic dynamics of the SIS model and its similarity to the HL model, in the presence of immigration. In Sec.~\ref{Sec:Sto_SIS}, the stochastic version of the model in the presence of multiplicative noise is considered and the universal features of the bifurcation diagram explored. In Sec.~\ref{Sec:sto_w_vac}, the effect of vaccination on the dynamics of the stochastic SIS model, in the absence of immigration, is studied. In Sec.~\ref{Sec:VH}, a vaccine-hesitancy model is proposed, with general applicability to infectious diseases that require vaccination, and the relevance of the model in the context of the COVID-19 vaccination program is pointed out. In Sec.~\ref{Sec:epi_model_vac}, a general form of the steady-state PDF of the basic reproduction number is derived based on a Beta-distributed fraction of the vaccinated population. Finally, Sec.~\ref{Sec:conclusion} contains concluding remarks.

\section{Deterministic Dynamics of SIS model}\label{Sec:det_dyn}
    The rate equations describing the dynamics of the SIS model with immigration are \cite{oregan2013}

    \begin{equation}\label{eq:x_SIS}
        \frac{dx}{dt}=-\beta xy-\eta x+\gamma y,
    \end{equation}

    \begin{equation} \label{eq:y_SIS}
        \frac{dy}{dt}=\beta xy+\eta x-\gamma y,
    \end{equation}

    \noindent where $x$ and $y$ denote, respectively, the fractions of the susceptible and infected populations in a total population of size $N_p$, with $x+y=1$. The parameter $\beta$ is the rate constant for the transmission of infection and the rate constant $\gamma$ is associated with the rate of transfer from the infected to the susceptible subpopulation. The model further assumes that infection occurs through contact with infected individuals trickling into the population at a small rate $\eta$, either by susceptible individuals transiently leaving the population, getting infected elsewhere, and reentering the population or through infected visitors briefly entering the population and infecting susceptible individuals [6]. This keeps the population size $N_p$ constant with the total infection rate given by $\beta y+\eta$. Since $x+y=1$, one can replace Eqs. (1) and (2) by a single rate equation

    \begin{equation} \label{eq:y_main}
        \frac{dy}{dt}=f(y)=\beta y(1-y)+\eta (1-y)-\gamma y.
    \end{equation}

    Two special cases of the model, which have so far been studied, are the SIS model with no immigration $(\eta=0)$ \cite{mamis2023} and the  HL model, with  $\eta=1/2$, $\eta+\gamma=1$ \cite{horsthemke1984,bose2022}. In the case of the SIS model with no immigration, Eq.~\pref{eq:y_main} has two steady-state solutions: 

    \begin{equation}
        y^{*}=0,~y^{*}=1-\frac{\gamma}{\beta}=1-\frac{1}{R_{0}}.
    \end{equation}

\begin{figure}
        \includegraphics[scale=0.62]{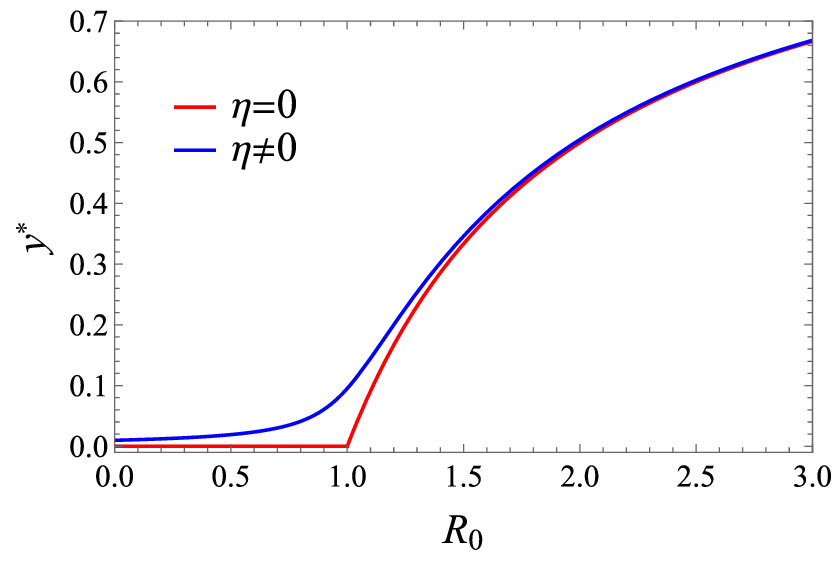}
        \caption{Steady-state value of the fraction of infected population ($y^*$) vs $R_0$, the basic reproduction number. In the absence of immigration, there is an absorbing phase transition at $R_0=1$  (red solid line). No such transition takes place for immigration rate $\eta\neq 0$ (blue solid line).}\label{fig:bifur_plot_deterministic}
    \end{figure}

    The parameter $R_{0}=\mfrac{\beta}{\gamma}$ is known as the basic reproduction number, a key parameter in epidemiology, defined to be the mean number of secondary infections caused by a single infected individual in a population of susceptible individuals. The model undergoes a transition from an “active” to an absorbing phase (Fig.~\ref{fig:bifur_plot_deterministic}, red solid line) at the transcritical bifurcation point $R_{0}=1$. In the active phase, the infected fraction of the population ($y^*$) is a finite quantity whereas it is zero in the absorbing phase, i.e., the infection is totally eliminated from the population. The absorbing states, characterized by the trapping of the dynamics with no possibility for escape, constitute a universal feature of several nonequilibrium statistical physics models \cite{pastorsatorras2015}.

    For  $\eta\neq 0$, there is a single steady-state solution $(\eta>0)$ given by   

    \begin{equation}
        y^{*}=\frac{(\beta-\eta-\gamma)+\sqrt{(\beta-\eta-\gamma)^{2}+4\beta \eta}}{2\beta}.
    \end{equation}

    There is now no absorbing phase transition at $R_{0}=1$ (Fig.~\ref{fig:bifur_plot_deterministic}) with $y^*$ never falling to zero but having a small value in the parameter regime $R_0<1$.  The return time $\tau$, defined to be the average time taken by the system to regain its steady state after being perturbed away from it, is given by the expression, $\tau=\mfrac{1}{\left|\lambda'\right|},~\lambda'=f^{'}(y)_{y=y^*}$ with $f(y)$ as given in Eq.~\pref{eq:y_main}. For $\eta=0$, the return time, $\tau=\mfrac{1}{\beta-\gamma}$, diverges (critical slowing down) at the transcritical bifurcation point $R_{0} =1$. For $\eta\neq 0$, the return time $\tau=\mfrac{1}{\sqrt{(\beta-\gamma-\eta)^{2}+4\beta\eta}}$ increases as $R_{0}\rightarrow 1$ but does not diverge.

\section{Stochastic SIS model with multiplicative noise}\label{Sec:Sto_SIS}

    In our model, we introduce multiplicative noise in the infection transmission parameter as $\beta \rightarrow \beta+\sigma \xi$ where $\xi$ represents the fluctuations, with intensity $\sigma^{2}$ around the deterministic value. We assume $\xi$ to have a Gaussian white noise form with mean zero and unit variance.  Substituting the expression for the fluctuating parameter $\beta$ in Eq.~\pref{eq:y_main}, one gets 


    \begin{equation}\label{eq:y_main_SDE}
        \frac{dy}{dt}=\beta y(1-y)+\eta (1-y)-\gamma y+y(1-y)\sigma \xi.
    \end{equation}

    The state variable $y$ is now a random variable and has thus a probability distribution. Let $p(y,t)dy$ be the probability of finding $y$ in the interval $(y,~y+dy)$ at time t. In the It\^o formalism, we get, from Eq.~\pref{eq:y_main_SDE}, the Fokker-Planck equation (FPE) \cite{horsthemke1984,bose2022}

    \begin{equation}\label{FP_eq}
        \frac{\partial p(y,t)}{\partial t}=-\frac{\partial}{\partial y}[f(y)p(y,t)]+\frac{\sigma^{2}}{2}\frac{\partial^{2}}{\partial y^{2}}[g^{2}(y)p(y,t)],
    \end{equation}

    \noindent where $f(y)=\beta y(1-y)+\eta (1-y)-\gamma y$ and $g(y)=y(1-y)$. The steady-state solution of the FPE is given by the expression

    \begin{equation}\label{pdf_formula}
        p_{s}(y)=\frac{\mathcal{N}}{g^{2}(y)}\exp{\left[\frac{2}{\sigma^2}\int^{y}\frac{f(q')}{g^{2}(q')}dq'\right]},
    \end{equation}

    \noindent where $\mathcal{N}$ is the normalization constant. The computation of the integral yields
    
    \begin{multline}\label{eq:pdf_gen}
        p_{s}(y)=\frac{\mathcal{N}}{y^{2}(1-y)^{2}}\exp \left[-\frac{2}{\sigma^{2}}\left\{\frac{\eta}{y}+\frac{\gamma}{1-y}+\right.\right. \\ 
        \left.\left. (\beta-\gamma+\eta)\ln{\left(\frac{1-y}{y}\right)}\right\}\right].
    \end{multline}

    The two special cases of the steady-state probability distribution correspond to the SIS model without immigration \cite{mamis2023} and the HL model of population genetics \cite{horsthemke1984,bose2022}.  

    Defining $R_{0}=\mfrac{\beta}{\gamma}$, $V=\mfrac{\sigma^2}{\beta}$ and $R_{1}=\mfrac{\eta}{\gamma}$,  Eq.~\pref{eq:pdf_gen} can be recast as

    \begin{multline}\label{eq:pdf_main}
        p_{s}(y)=\mathcal{N}y^{\frac{2}{V}\left(1-V-\frac{1}{R_{0}}+\frac{R_{1}}{R_{0}}\right)}(1-y)^{-\frac{2}{V}\left(1+V-\frac{1}{R_{0}}+\frac{R_{1}}{R_{0}}\right)} \\
        \times\exp{\left[-\frac{2}{VR_{0}}\left(\frac{1}{1-y}+\frac{R_{1}}{y}\right)\right]}.
    \end{multline}

    The probability density function (PDF), $p_{s} (y)$, is integrable for all parameter values with $\eta\neq 0$. The integrability condition, $\int_{0}^{1}p_{s}(y)dy=1$, in the case of $\eta=0$ is fulfilled only if $V<2\left(1-\mfrac{1}{R_{0}}\right)$ \cite{mamis2023}. 

    The extrema $y_{m}$, $\left(\mfrac{dp_{s}}{dy}\bigg|_{y_m}\equiv F(y_m)=0\right)$, of the steady-state PDF are obtained as the roots of the equation

    \begin{multline}\label{eq:Fym}
        F(y_{m})=R_{1}-(R_{1}+1)y_{m}+R_{0}y_{m}(1-y_{m})- \\
        R_{0}Vy_{m}(1-y_{m})(1-2y_{m})=0.
    \end{multline}

    Equation~\pref{eq:Fym} has the form of a steady-state equation with the extremum $y_{m}$ serving as the “state” variable. Note that in the absence of noise, i.e., $V=0$,  Eq.~\pref{eq:Fym} has the form of the deterministic steady-state equation $f(y)=0$ (Eq.~\pref{eq:y_main}). The significance of the extrema arises from the fact that they are the appropriate indicators of a transition from one stochastic dynamic regime to another. For example, the number of maxima of $p_{s}(y)$ changes from one to two in the case of a transition from unimodality to bimodality. The term noise-induced transition implies that in the deterministic case, i.e., in the absence of noise, there is no analogous transition (bifurcation) from one to two stable steady states. Since the maxima represent the most probable values preferentially accessed in experiments, their number may be regarded as representative of a “phase”. By now, it is well established that in contrast to additive noise, which does not depend on state variables, multiplicative noise can bring about a noise-induced transition when the noise intensity $\sigma^{2}$ is sufficiently large. In our SIS model, there is a single stable steady state defining a single phase in the deterministic case whereas in the stochastic case, the bifurcation diagram of the PDF, shown in Fig.~\ref{fig:bifur_im_coex} (discussed again in detail later in this section), has two distinct phases corresponding to unimodal and bimodal steady-state probability distributions.

    In the bifurcation diagram (Fig.~\ref{fig:bifur_im_coex}), the boundaries between the two phases, shown by solid lines end at a cusp point. Across each boundary, a saddle-node (SN) bifurcation takes place, i.e., the number of solutions of Eq.~\pref{eq:Fym} changes from three to one. The conditions for the occurrence of the SN bifurcation are $F(y_{m})=0$ (Eq.~\pref{eq:Fym}) and 

    \begin{multline}\label{eq:F1ym}
        F^{'}(y_{m})=-(R_{1}+1)+R_{0}(1-2y_{m})- \\
        R_{0}V(1-6y_{m}+6y_{m}^{2})=0.
    \end{multline}

    From Eqs.~\pref{eq:Fym} and ~\pref{eq:F1ym}, one can derive the two parametric equations, defining the boundaries in the bifurcation diagram, as

    \begin{equation}\label{R_0_eq}
        R_{0}=\frac{R_{1}(1-6y_{m}+6y_{m}^{2})-(R_{1}+1)y_{m}(4y_{m}^{2}-3y_{m})}{2y^{2}_{m}(1-y_{m})^{2}},
    \end{equation}

    \begin{equation}\label{V_eq}
        V=\frac{R_{1}(1-2y_{m})+(R_{1}+1)y^{2}_{m}}{R_{1}(1-6y_{m}+6y_{m}^{2})-(R_{1}+1)y_{m}(4y_{m}^{2}-3y_{m})}.
    \end{equation}

    Figure~\ref{fig:bifur_im_coex} shows the $V$ versus $R_{0}$ bifurcation diagram with $R_{1}=0.01$. The diagram shows the existence of two distinct phases, unimodal and bimodal, with the number of peaks (maxima) of the steady-state PDF (Eq.~\pref{eq:pdf_main}) being one and two, respectively, in the two phases. The boundaries between the two regions are lines of SN bifurcation that terminate at a cusp point, marked by a solid blue dot. The behavior is reminiscent of a line of first-order transitions terminating at a critical point in the phase diagrams of thermodynamic phase transitions like liquid-gas and paramagnetic to ferromagnetic transitions. Fig.~\ref{fig:multiple_pdfs_im} shows the steady-state PDFs, both unimodal and bimodal, in different regions of the bifurcation diagram. At the cusp point (critical point), the following conditions are satisfied:

    \begin{figure}
        \centering
        \includegraphics[scale=0.62]{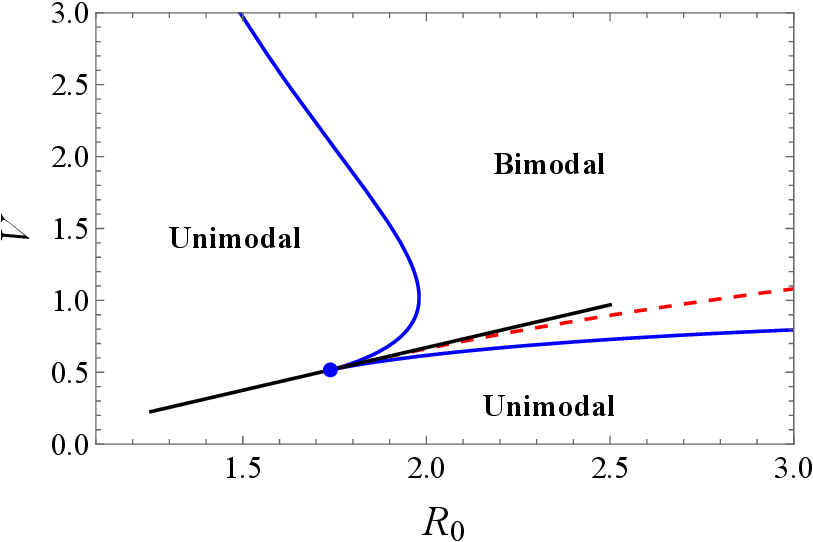}
        \caption{Bifurcation diagram ($V$ versus $R_0$) in the $R_0-V$ plane for the SIS model with immigration $\left(R_1=\mfrac{\eta}{\gamma}=0.01\right)$ $\left(R_1=\mfrac{\eta}{\gamma}=0.01\right)$. The diagram shows two distinct regions of unimodality and bimodality. The solid blue dot is the critical point, the solid black line is the tangential line entering the region of bimodality at the critical point, and the red dashed line represents the coexistence line.}\label{fig:bifur_im_coex}
    \end{figure}


    \begin{figure*}
    \centering
        \begin{subfigure}{.3\textwidth}
            \centering
            \includegraphics[scale=0.39]{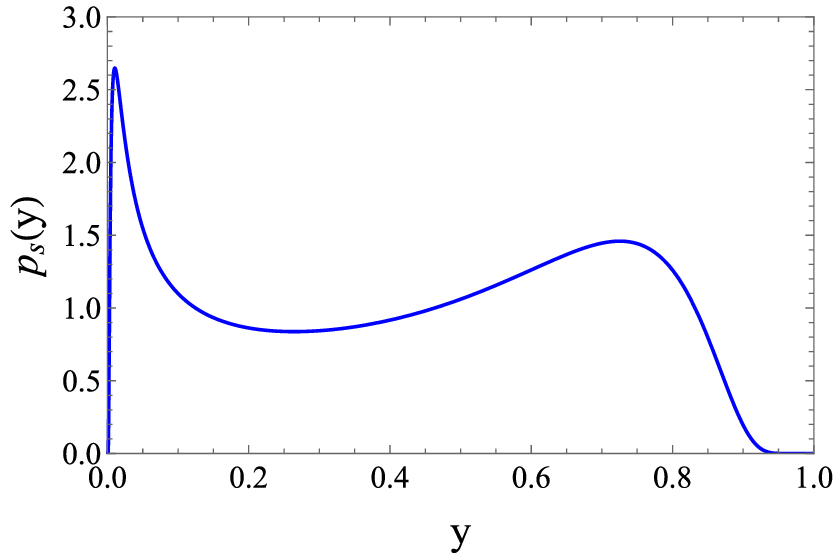}
            \caption{}
            \label{}
        \end{subfigure}
        \begin{subfigure}{.3\textwidth}
            \centering
            \includegraphics[scale=0.39]{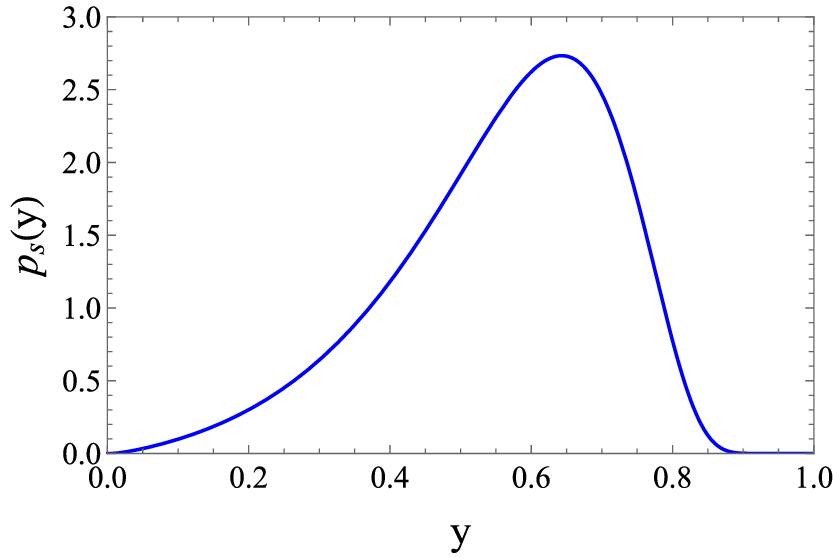}
            \caption{}
            \label{}
        \end{subfigure}
        \begin{subfigure}{.3\textwidth}
            \centering
            \includegraphics[scale=0.39]{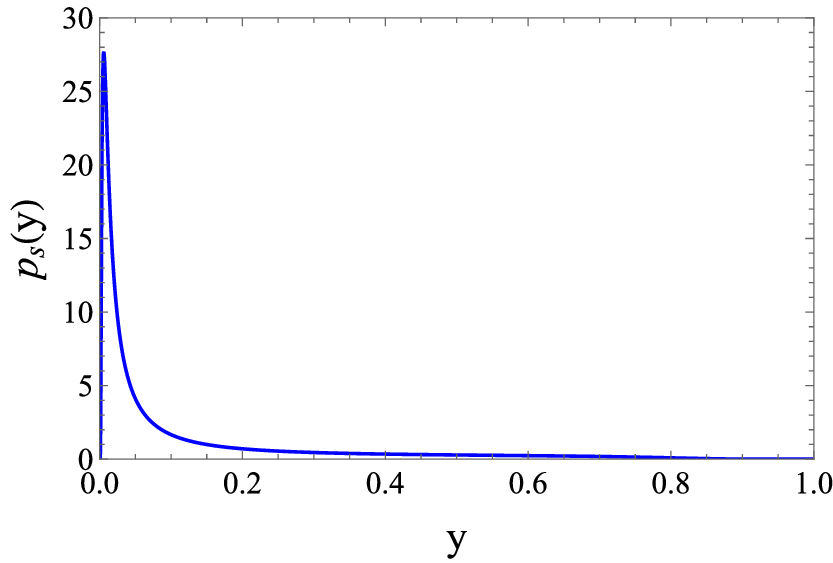}
            \caption{}
            \label{}
        \end{subfigure}
        \caption{Steady-state PDF $p_s(y)$ versus  $y$ (Eq.~\pref{eq:pdf_main}) for $R_1=0.01$. The figure on the left (a) shows a bimodal PDF and the other two figures, (b) and (c), represent PDFs in the region of unimodality, as shown in the bifurcation diagram of Fig.~\ref{fig:bifur_im_coex}. The sets of $(R_{0},~V)$ values in (a), (b), and (c) are (2.5, 1.0), (2.5, 0.4), and (1.5, 1.5), respectively.}
        \label{fig:multiple_pdfs_im}
    \end{figure*}


    \begin{equation}\label{eq:crit_rel}
        F(y_{m})=0,~F^{'}(y_{m})=0,~F{''}(y_{m})=0.
    \end{equation}

    
    The third condition in Eq.~\pref{eq:crit_rel}, coming from Eq.~\pref{eq:F1ym}, yields the expression $y_{m}=\mfrac{1}{2}-\mfrac{1}{6V}$. Substituting the expression for $y_{m}$ in Eqs.~\pref{eq:Fym} and \pref{eq:F1ym}, we get

    \begin{equation}\label{eq:cusp_rel_1}
        R_{0}=\frac{6V(R_{1}+1)}{3V^2+1}  
    \end{equation}

    and

    \begin{equation}\label{eq:cusp_rel_2}
        R_{0}=\frac{27V^{2} (1-R_{1} )-9V(R_{1}+1)}{9V^{2}-1}.
    \end{equation}

\begin{figure}
        \centering
        \includegraphics[scale=0.62]{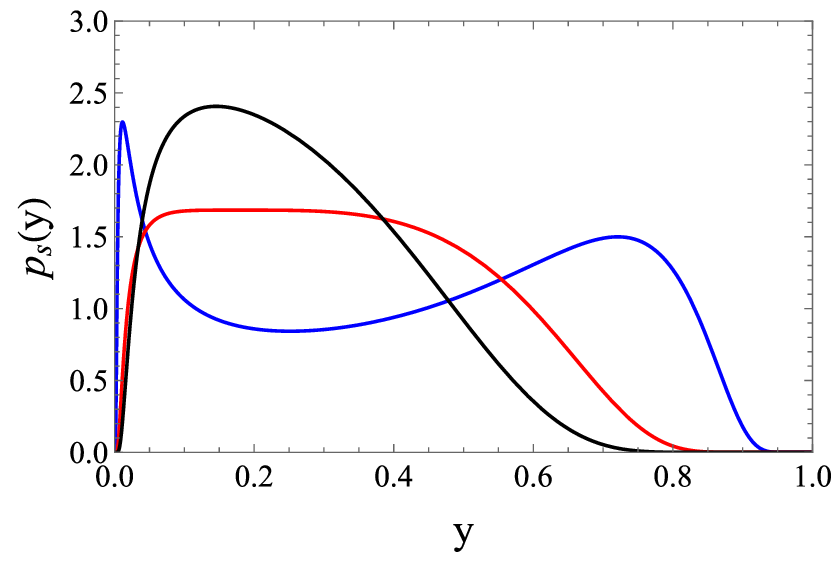}
        \caption{The steady-state PDFs, $p_s (y)$ versus $y$, along the tangential line of Fig.~\ref{fig:bifur_im_coex}. The figure illustrates the noise-induced transition from unimodality to bimodality as the noise intensity is increased. The parameter values are $R_{0}=2.5,~1.73854,~\text{and}~1.5$ for the bimodal (blue solid line), critical (red solid line), and unimodal (black solid line) PDFs, respectively.}\label{fig:pdfs_superimpose}
    \end{figure}

    From  Eqs.~\pref{eq:cusp_rel_1} and \pref{eq:cusp_rel_2}, the critical point for $R_{1}=0.01$ is computed as $(R_{0},V)=(1.73854,0.516404)$. In the bifurcation diagram, the black solid line represents the tangential line, entering the region of bimodality, at the critical point and the red dashed line is the coexistence line. The tangential line is given by the straight line equation $V=0.594065R_{0}-0.516402$. The steady-state PDFs, $p_{s} (y)$ versus $y$, along the tangential line are shown in Fig.~\ref{fig:pdfs_superimpose} illustrating noise-induced transition from unimodality to bimodality. The PDF has a characteristic flat top at the critical point. 

\begin{figure}
        \centering
        \includegraphics[scale=0.62]{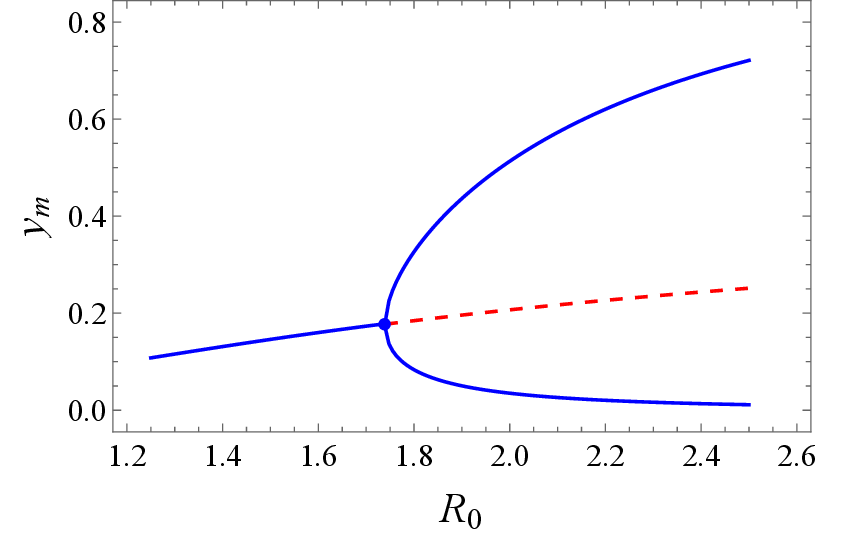}
        \caption{Plot of $y_m$ (Eq.~\pref{eq:Fym}) versus $R_0$ showing a pitchfork bifurcation at the critical (cusp) point with the tangential line serving as the axis of the bifurcation plot. The blue solid and red dashed lines represent stable and unstable states of $y_m$, respectively. }\label{fig:sp_bifur}
    \end{figure}

\begin{figure}
        \centering
        \includegraphics[scale=0.62]{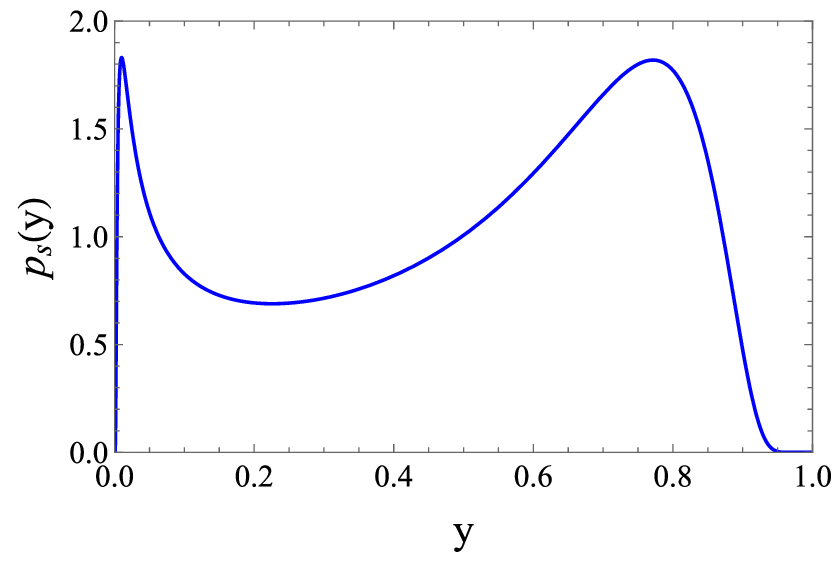}
        \caption{A bimodal PDF at $(R_0,V)=(2.812,1.015)$ on the coexistence line of Fig.~\ref{fig:bifur_im_coex} showing that the peaks are of the same height.}\label{fig:pdf_coex}
    \end{figure}

    In Fig.~\ref{fig:pdfs_superimpose}, the values of $R_{0}$ are $R_{0}=2.5,~1.73854~\text{(critical point),}~\text{and}~1.5$ for the bimodal (blue solid line), critical (red solid line), and unimodal (black solid line) PDFs, respectively, with the corresponding values of $V$ obtained from the straight line equation for the tangential line. The cusp point coincides with a supercritical pitchfork (SP) bifurcation point with the tangential line serving as the axis of the bifurcation plot and the steady-state values of $y_{m}$ determined as roots of the polynomial equation in Eq.~\pref{eq:Fym} (Fig.~\ref{fig:sp_bifur}). For a specific value of $R_{0}$, the magnitude of  $V$  is obtained from the straight line equation describing the tangential line. The red dashed line (Fig.~\ref{fig:bifur_im_coex}) is the coexistence line along which the peaks of the bimodal PDF are of the same height. Figure~\ref{fig:pdf_coex} shows such a bimodal PDF with $(R_{0}, V)=(2.812,1.015)$. 

    Following the procedure outlined in \cite{bose2022criticality}, we show that the critical point transition in the stochastic SIS model with immigration belongs to the mean-field Ising universality class. The mean-field equation of state of the Ising model close to the critical point is given by \cite{bose2022}

    \begin{equation}\label{eq:MF_Ising}
        h-\theta m-\frac{m^{3}}{3}=0,
    \end{equation}

     \noindent where $m$ is the average magnetization per spin, $h$ the reduced magnetic field, and $\theta=\left(\mfrac{T-T_{c}}{T_{c}}\right)$  is the reduced temperature with $T_{c}$ being the critical transition temperature. The critical point is given by $h=0$, $\theta=0$. The order parameter is the spontaneous magnetization, $m(h=0)$, which has a nonzero (zero) value below (above) $T_{c}$.

    The expression $F(y_{m})=0$ in Eq.~\pref{eq:Fym} is Taylor expanded around a point $y_{c}$ to the third order with the choice of  $y_{c}$ dictated by the condition   $F^{''} (y_{c})=0$. The resulting equation has a form similar to the magnetic equation of state (Eq.~\pref{eq:MF_Ising}) since a second-order term is absent in both the expressions. On Taylor expansion, we get

    \begin{equation}\label{eq:taylor_exp}
        F(y_{c})+F^{'} (y_{c})(y-y_{c})+F^{'''} (y-y_{c}) \frac{(y-y_{c})^{3}}{3!}=0. 
    \end{equation}

    Comparing with Eq.~\pref{eq:MF_Ising}, one gets the following relations between the thermodynamic and dynamic quantities:

    \begin{equation}\label{eq:comparison}
        m=\frac{(y_{m}-y_{c} )}{y_{c}},~h=\frac{-2F(y_{c})}{F^{'''}(y_{c})y_{c}^{3}},~\theta=\frac{2F^{'}(y_{c})}{F^{'''}(y_{c}) y_{c}^{2}}.   
    \end{equation}

    The special point $y_{c}$ is given by the expression

    \begin{equation}\label{eq:crit_point}
        y_{c}=\frac{1}{2}-\frac{1}{6V}=\frac{1}{2}-\frac{\beta}{6\sigma^{2}}.
    \end{equation}

    At the critical point, $m=0$, i.e., $y_{m}=y_{c}$. Also, since both $F(y_{c})$ and $F^{'}(y_{c})=0$ at this point (putting $y_{m}=y_{c}$ in Eq.~\pref{eq:crit_rel}), we get $h=0$, $\theta=0$. The critical point is also the SP bifurcation point. From Eqs.~\pref{eq:taylor_exp} – \pref{eq:crit_point}, one finds that the critical point transition at the cusp point in the bifurcation diagram of the SIS model belongs to the mean-field Ising universality class. The thermodynamic quantities of interest and their scaling relations close to the critical point are \cite{horsthemke1984,bose2022,erez2019,bose2019}

    \begin{equation}\label{eq:therm1}
        m(h=0)\sim\pm(-3\theta)^{\beta_{c}},~\beta_{c}=\frac{1}{2},~\theta<0,
    \end{equation}

    \begin{multline}\label{eq:therm2}
        \chi=\left(\frac{\partial m}{\partial h}\right)_{h=0}\sim \theta^{-\gamma_{c}}~(\theta>0),~\chi\sim(-2\theta)^{\gamma_{c}}~(\theta<0), \\ \gamma_{c}=1, 
    \end{multline}

    \begin{equation}\label{eq:therm3}
        m=(3h)^{\frac{1}{\delta}},~\theta=0,~\delta=3. 
    \end{equation}

\begin{figure}
        \centering
        \includegraphics[scale=0.62]{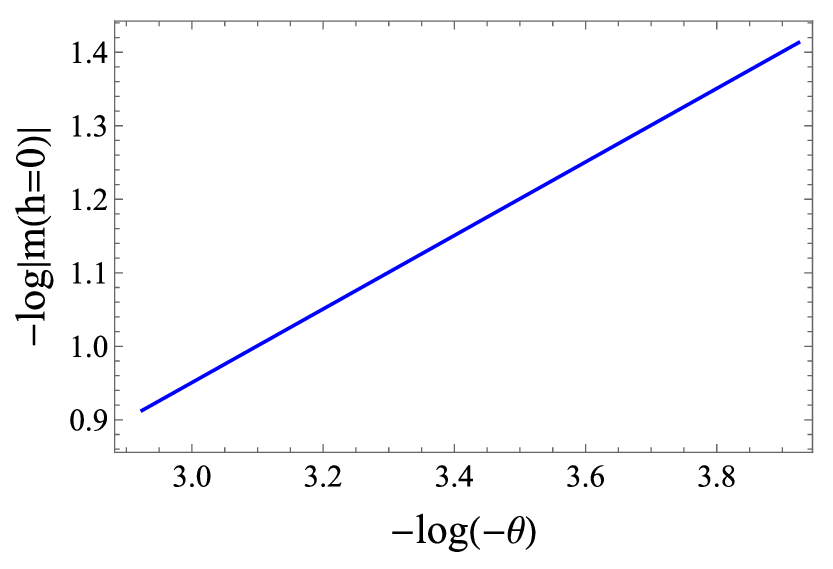}
        \caption{Plot of $-\log\left|m(h=0)\right|$ versus $-\log(-\theta)$ calculated directly from Eq.~\pref{eq:comparison} with the reduced field $h=0$. The slope of the straight line yields the Ising mean-field exponent $\beta_c=1/2$ for the order parameter.}\label{fig:ising_fig}
    \end{figure}

    The thermodynamic quantity $\chi$ represents the dimensionless susceptibility, the response function of a magnetic system. In the mapping between the thermodynamic and dynamic models, the temperature $T$ is analogous to the noise intensity $\sigma^{2}$ (incorporated in the parameter $V=\mfrac{\sigma^{2}}{\beta}$ in our model).  Figure~\ref{fig:ising_fig} shows a plot of $-\log\left|m(h=0)\right|$ versus $-\log(-\theta)$, calculated directly from the mapping relations given in Eq.~\pref{eq:comparison}, with the reduced field $h=0$ $(F(y_{c})=0)$. The slope of the straight line is determined as $\beta_{c}=\mfrac{1}{2}$ in agreement with Eq.~\pref{eq:therm1}.

\section{Stochasticity with vaccination}\label{Sec:sto_w_vac}

    In this section, we include the effect of vaccination on the control and eradication of the infection in the stochastic SIS model. In the deterministic case, the vaccination parameter $q$, defined as the fraction of the vaccinated population, is introduced in the dynamical equation (Eq.~\pref{eq:y_main}), as (see \cite{oregan2013})

    \begin{equation} 
        \frac{dy}{dt}=\beta (1-q) y(1-y)+\eta (1-y)-\gamma y.
    \end{equation}

    To facilitate the analysis of the effect of vaccination, we consider the case $\eta=0$, i.e., there is no immigration. Defining an effective reproduction number $R=R_0 (1-q)$, the transcritical bifurcation  occurs at $R=1$, i.e., at a value of the basic reproduction number $R_0>1$ $\left(q=1-\mfrac{R}{R_0}\right)$. Thus, vaccination, as is to be expected, has a favorable effect on the eradication of the infection. The critical vaccination coverage, $q_c$, is defined to be the value of $q$ at which  the effective reproduction number,  $R=1$, i.e., $q_c=1-\mfrac{1}{R_0}$. If, for example, $R_0=10$, $q_c=0.9$ and at least 90\% of the population has to be vaccinated for the total eradication of the infection. 

\begin{figure}
        \centering
        \includegraphics[scale=0.62]{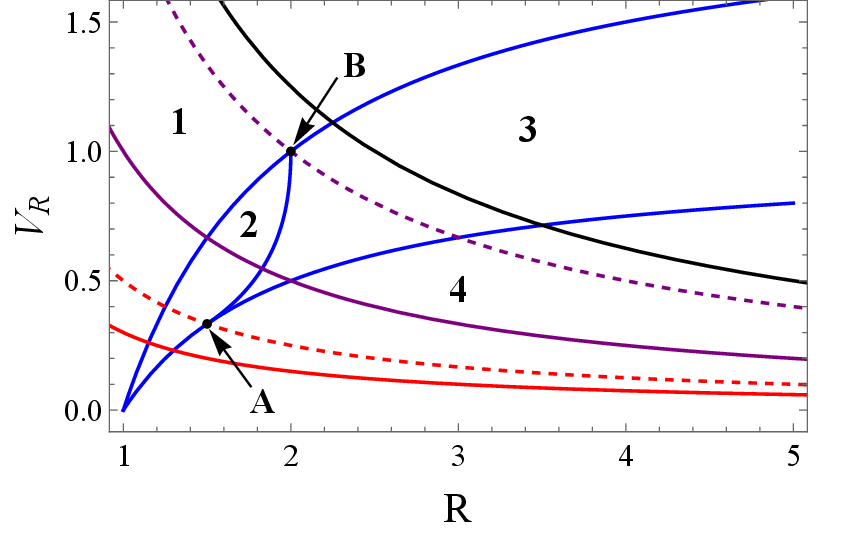}
        \caption{Plots of $RV_R=\mfrac{\sigma^2}{\gamma}$ drawn on the bifurcation ($V_R$ versus $R$) diagram for $\eta=0$. The values of $\mfrac{\sigma^2}{\gamma}$ for the red, purple, and black solid lines are 0.3, 1.0, and 2.5, respectively. Red and Purple dashed lines pass through the intersection points, A and B,  of the bifurcation lines, with $\mfrac{\sigma^2}{\gamma}$= 0.5 and 2.0, respectively. }\label{fig:sto_bifurcation_RVlines}
    \end{figure}

    In the case of the stochastic SIS model, which includes vaccination and with $\eta=0$, the steady-state  PDF is obtained from Eq.~\pref{eq:pdf_main} by putting $R_1=0$ and replacing $\beta$ by $\beta(1-q)$ in the definitions of $V$ and $R_0$. Following the procedure outlined in Sec.~\ref{Sec:Sto_SIS}, the bifurcation diagram (Fig.~\ref{fig:sto_bifurcation_RVlines}) is obtained in the  $R$-$V_R$  plane with $R=R_0 (1-q)$, $R_0=\mfrac{\beta}{\gamma}$  and   $V_R=\mfrac{V}{(1-q)}$, $V=\mfrac{\sigma^2}{\beta}$. The same bifurcation diagram has been obtained earlier \cite{mamis2023} in the $R_0-V$ plane, i.e., in the absence of vaccination. There are four distinct regions in the bifurcation diagram labeled ``1-4" with the bifurcation lines marked in blue. The steady-state PDFs in the different regions are a delta function at $y=0$ (region 1), unimodal with maximum at $y=0$ (region 2), bimodal with one maximum at $y=0$  (region 3), and unimodal with maximum at nonzero $y$ \cite{mamis2023}. To analyze the combined effect of stochasticity and vaccination, we draw $RV_R=\mfrac{\sigma^2}{\gamma}=\text{constant}$ lines on the bifurcation diagram in Fig.~\ref{fig:sto_bifurcation_RVlines}. Three such lines, marked with red, purple, and black solid lines, for  $\mfrac{\sigma^2}{\gamma}$= 0.3, 1.0, and 2.5, respectively, are shown in the figure. The lines passing through the intersection points A and B of the bifurcation lines correspond to  $\mfrac{\sigma^2}{\gamma}$= 1/2 and 2. If  $\mfrac{\sigma^2}{\gamma}<1/2$, the $RV_R=\text{constant}$ lines pass through the regions of the bifurcation diagram in the order $4\rightarrow 2\rightarrow 1$. If  $1/2<\mfrac{\sigma^2}{\gamma}<2$, the order of traversal is $4\rightarrow 3\rightarrow 2\rightarrow 1$ and for $\mfrac{\sigma^2}{\gamma}>2$, the order is $4\rightarrow 3\rightarrow 1$. Over the full range of values of  $\mfrac{\sigma^2}{\gamma}$, the $RV_R= \text{constant}$ lines eventually  cross the upper bifurcation line $V_R=2\left(1-\mfrac{1}{R}\right)$ into the region ``1" as $R$ is decreased. At the crossing point, one obtains the general relation

    \begin{equation}\label{eq:critical_rep_no}
       R_{c}=1+\frac{\sigma^2}{2\gamma}
    \end{equation}

    \noindent since  $RV_R=\mfrac{\sigma^2}{\gamma}$. The same relationship holds true in the absence of vaccination.
   
\begin{figure*}
	\centering
	  \begin{subfigure}{.45\textwidth}
	      \centering
	      \includegraphics[scale=0.55]{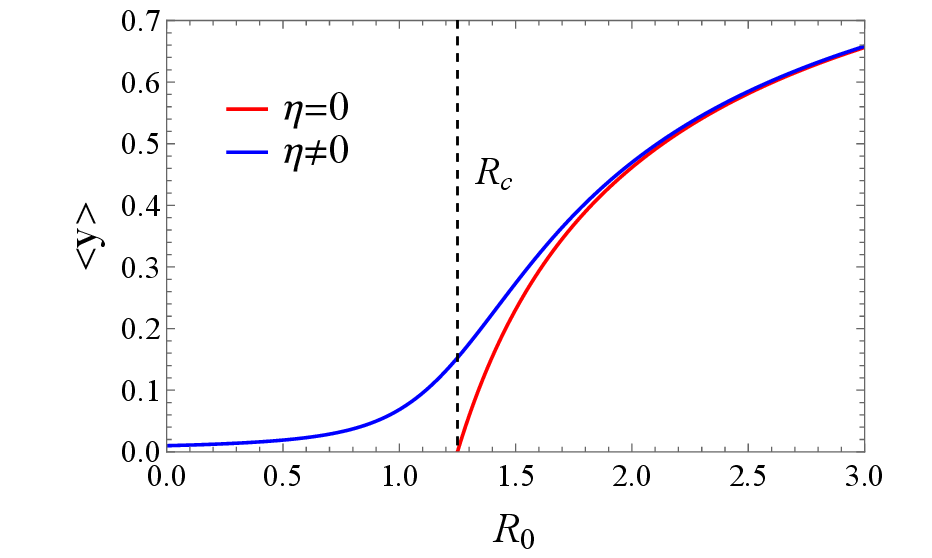}
	      \caption{}
	      \label{fig:mean}
	  \end{subfigure}
	  \hspace{20pt}
	  \begin{subfigure}{.45\textwidth}
	      \centering
	      \includegraphics[scale=0.55]{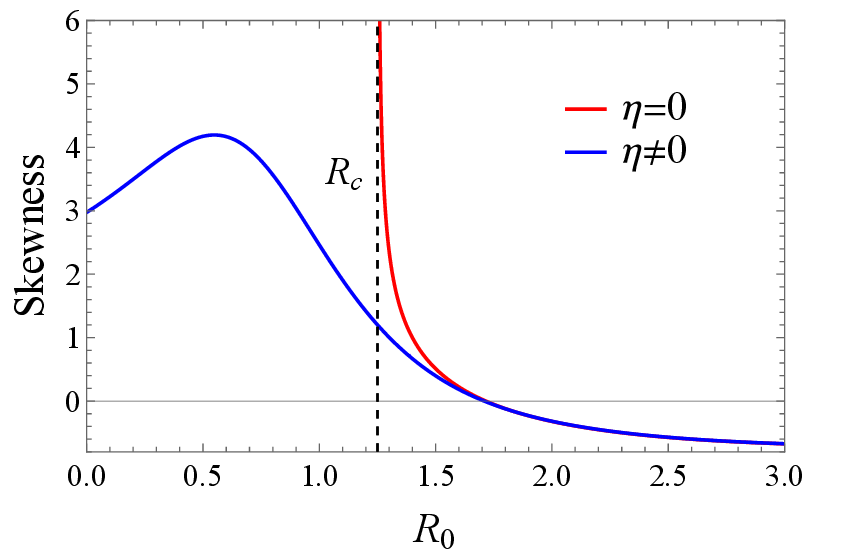}
	      \caption{}
	      \label{fig:skew}
	  \end{subfigure}
	  \begin{subfigure}{.45\textwidth}
	      \centering
	      \includegraphics[scale=0.55]{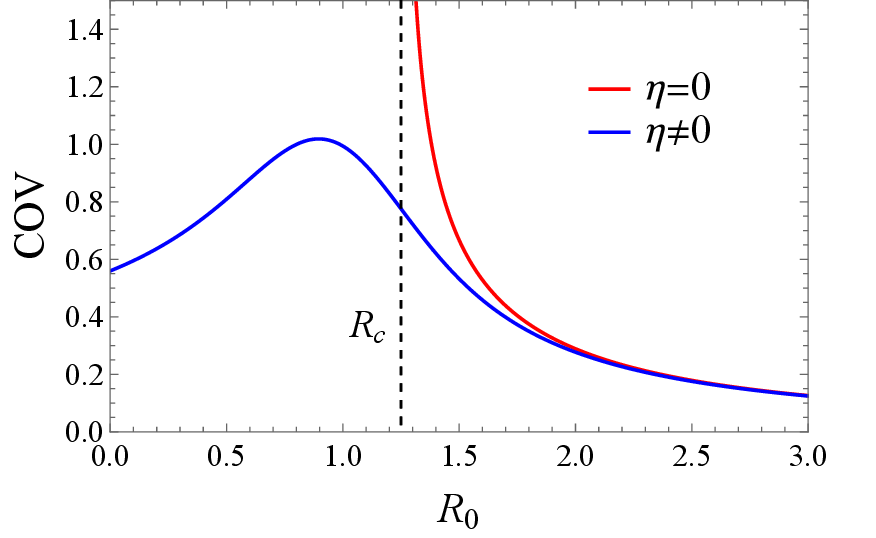}
	      \caption{}
	      \label{fig:covar}
	  \end{subfigure}
	  \begin{subfigure}{.45\textwidth}
	      \centering
	      \includegraphics[scale=0.55]{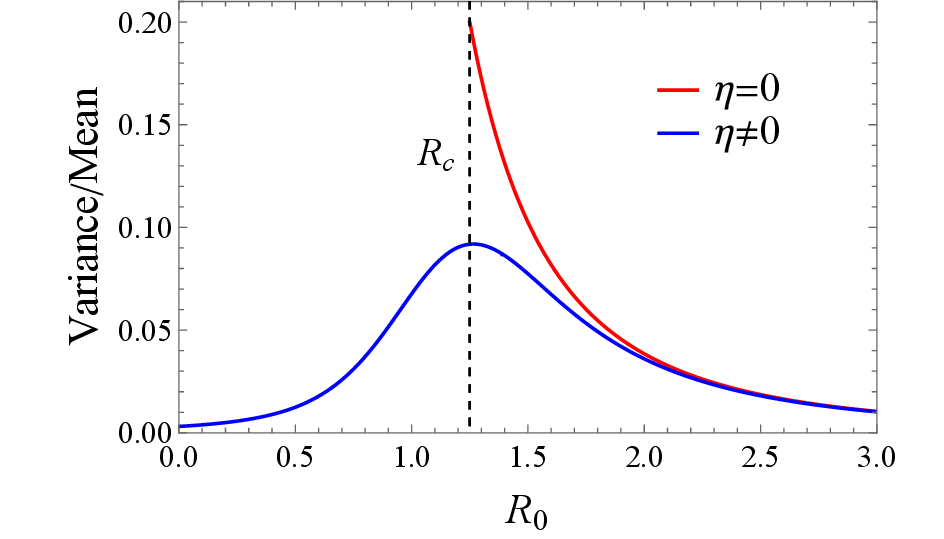}
	      \caption{}
	      \label{fig:ratio}
	  \end{subfigure}
	  \caption{(a) Mean, (b) skewness, (c) coefficient of variation (COV), (d) Variance/Mean of steady-state PDF for $\eta=0$ (red solid line) and $\eta\neq 0$ (blue solid line) versus $R_0$. The parameter values are $R_1=0.01$, $\mfrac{\sigma^2}{\gamma}=0.5$.}
	  \label{fig:stat_parameters}
	\end{figure*}

    If the effective reproduction number $R$ is less than $R_c$, the system is in region ``1" in which, because of the delta function form of the PDF at $y=0$, the infection is totally eradicated. Thus, the disease-free condition in the epidemic model is set by the threshold value $R_c$ of the effective reproduction number. The effect of stochasticity is to promote the total eradication of the infection since the threshold for disease extinction rises from $R=1$ in the deterministic case to $R_{c}=1+\mfrac{\sigma^2}{2\gamma}$ when stochasticity is included. Accordingly, the critical vaccination coverage is given by  $q_c=1-\mfrac{1}{R_0}$  in the deterministic case and $q_c=1-\mfrac{1}{R_0} -\mfrac{\sigma^2}{2R_0\gamma}$ in the stochastic case. The combination of stochasticity and vaccination thus offers the advantage of full disease eradication with a lower fraction of the vaccinated population. A result similar to that shown in Eq.~\pref{eq:critical_rep_no} forms the content of a theorem proposed and proved in the case of the stochastic SIS model without vaccination \cite{gray2011}. Below the threshold value set by $R_c$ (Eq.~\pref{eq:critical_rep_no}), the infected fraction of the population in the steady state is zero with probability 1. As in the deterministic case, an absorbing phase transition takes place at the critical point set now by the threshold value $R_c$. Models of epidemic spreading share universal features with other models of nonequilibrium statistical physics exhibiting an absorbing phase transition \cite{pastorsatorras2015}. Such a transition is analogous to a critical point transition with the relevant order parameter $\rho$ going continuously to zero in a power-law fashion, $\rho\sim\left(\lambda-\lambda_c\right)^{\beta_c}$, as a parameter $\lambda$ approaches its critical value $\lambda_c$ with $\beta_c$ representing the critical exponent. The critical point $\lambda_c$ separates two distinct phases: $\rho>0$  for   $\lambda>\lambda_c$  and $\rho=0$ (absorbing phase)  for  $\lambda<\lambda_c$. In the case of the stochastic SIS model without immigration, the order parameter is given by the fraction of infected population and the relevant parameter is the basic reproduction number $R_0$ (or the effective reproduction number $R$ if vaccination is included). The order parameter is given by the mean value $\langle y\rangle$ of the fraction of infected population in the steady state with  $\langle y\rangle\sim\left(R_0-R_c\right)^{\beta_c}$. Figure~\sref{fig:stat_parameters}{fig:mean} shows the plot of $\langle y\rangle$ versus $R_0$ for both $\eta=0$ (red solid line) and $\eta\neq 0$ (blue solid line). In the first case, $\langle y\rangle\rightarrow 0$ with the critical exponent $\beta_c=1$, which is the same as the order parameter exponent in the case of an absorbing phase transition \cite{pastorsatorras2015}. For $\eta\neq 0$, the transition (Eq.~\pref{eq:therm1}) belongs to the mean-field Ising universality class with $\beta_c=\mfrac{1}{2}$.

    We further point out that even for $R>R_c$, the infected fraction of the population can fall to zero with a certain probability. This is because, in this parameter regime, the steady-state PDF has a $y=0$ maximum in both regions 2 and 3. As Fig.~\ref{fig:sto_bifurcation_RVlines} shows, the $R-V_R$ lines, depending on the associated  $\mfrac{\sigma^2}{\gamma}$  values, pass through either region 2 or 3 singly (red or black solid line) or both the regions (purple solid line). Thus, when $R>R_c$, the probability for the eradication of infection is finite for the whole range of  $\mfrac{\sigma^2}{\gamma}$ values. In the presence of immigration at a small rate $\eta$, a total eradication of the infection is not possible, as unlike in the $\eta=0$ case, the steady-state PDF  does not have a maximum with zero value in any part of the bifurcation diagram (Fig.~\ref{fig:bifur_im_coex}).

    Figures~\ref{fig:stat_parameters} (a) – (d) show the plots, respectively, of the mean infected fraction of the population in the steady state,  the skewness, coefficient of variation (COV), and the variance/mean ratio of the steady-state PDF, versus the basic reproduction number $R_0$, for both $\eta=0$ (red solid line) and $\eta\neq 0$ (blue solid line). The parameter values are fixed at $R_1=0.01$, $\mfrac{\sigma^2}{\gamma}=0.5$. The values of $V$ are determined from the relation $V=\mfrac{\sigma^2}{\gamma}\mfrac{1}{R_{0}}$ for different values of $R_{0}$. In the case of $\eta=0$, as the critical point, $R_c=1+\mfrac{\sigma^2}{2\gamma}=1.25$, of the absorbing phase transition is approached, the mean infected fraction falls to zero and the quantities associated with the steady-state PDF attain their maximal values. All four quantities provide early signatures of the approach to the critical point.  The corresponding plots for the $\eta\neq 0$ case are also shown to facilitate comparison. There is no absorbing phase transition and the plots in the last three cases reach a maximum value and then start to fall. In both cases, the skewness of the steady-state PDF changes from a negative to a positive value as the magnitude of $R_0$ decreases.

\section{Vaccine-hesitancy model and vaccination}\label{Sec:VH}

    In this section, we propose a VH \cite{franceschi2023} model, taking cues from Kirman’s ant model \cite{kirman1993,alfarano2005,kononovicius2018} to study the effect of VH on the spread of an epidemic. The Kirman model studies the dynamics of an ant colony, collecting food from two identical food sources located in the vicinity of the colony. Both a herding instinct and individual propensity are at work in the choice of a specific food source by an ant. The VH model is developed along the lines of the ant model with people in a community replacing ants in a colony. The choice to get vaccinated or not is equivalent to the choice of a particular food source by an ant. In the VH model, the population is divided into two categories: vaccine willing (Category A) and vaccine hesitant (Category B). The transition from one category to the other is stochastic in nature based on several factors like awareness about vaccination, rumors, and news on the adverse effect of vaccination. Let $N_{p}$ be the total number of individuals in the community with the number of individuals in Category A and Category B given by X  and $N_{p}-X$, respectively. As in the case of the ant model, the transition probabilities are

    \begin{equation}\label{transition_1}
        p(X\rightarrow X+1)=(N_{p}-X)\left\{\sigma_{1}+hX\right\}\Delta t,
    \end{equation}

    \begin{equation}\label{transition_2}
        p(X\rightarrow X-1)=X\left\{\sigma_{2}+h(N_{p}-X)\right\}\Delta t.
    \end{equation}

    The parameter $\sigma_{i}$ describes the individual propensity to change the state while the parameter $h$ is a measure of the herding tendency in bringing about a transition. The short time step $\Delta t$ ensures that only one stochastic event, $X\rightarrow X\pm 1,~X$, takes place in the small interval of time. We define two rescaled parameters as $\epsilon_{i}=\mfrac{\sigma_{i}}{h}$ and the dynamical variable as $x=\mfrac{X}{N_{p}}$, the fraction of vaccine-willing individuals in the total population. The time $t$ is also rescaled as $t_{s}=ht$. Due to the conservation of the population size, the fraction of vaccine-hesitant individuals is given by $1-x$. In the continuum limit $N_{p}\rightarrow\infty$, one can approximate the stochastic time evolution of the probability density $P(x,t_{s})$

    \begin{multline}
        \frac{\partial P\left(x,t_{s}\right)}{\partial t_{s}}=-\frac{\partial}{\partial x}\left[\left(\epsilon_{1}(1-x)-\epsilon_{2}x\right)P(x,t_{s})\right]+ \\ \frac{\partial^{2}}{\partial x^{2}}\left[x(1-x)P(x,t_{s})\right].
    \end{multline}

    The steady-state solution of the FPE is given by the well-known Beta distribution:

    \begin{equation}\label{eq:kirman_pst}
        p_{\text{st}}(x)=\frac{\Gamma(\epsilon_{1}+\epsilon_{2})}{\Gamma(\epsilon_{1})\Gamma(\epsilon_{2})}x^{\epsilon_{1}-1}(1-x)^{\epsilon_{2}-1},
    \end{equation}

    \noindent where $\Gamma$ denotes the Gamma function. The Beta distribution is well-known for its flexibility and models random variables with values confined to the unit interval (0, 1). The PDF assumes a variety of shapes which include the U shaped $\left(\epsilon_{1},\epsilon_{2}<1\right)$, unimodal symmetric $\left(\epsilon_{1}=\epsilon_{2}>1\right)$, unimodal asymmetric $\left(\epsilon_{1}\neq\epsilon_{2}\right)$, and uniform density $\left(\epsilon_{1}=1,\epsilon_{2}=1\right)$. In vaccine-related research, e.g., in the determination of the vaccine efficacy distribution, a standard technique in use is that of Bayesian inference for which the Beta distribution is a popular candidate \cite{gray2011}. 

\begin{figure*}
    \centering
        \begin{subfigure}{.45\textwidth}
            \centering
            \includegraphics[scale=0.55]{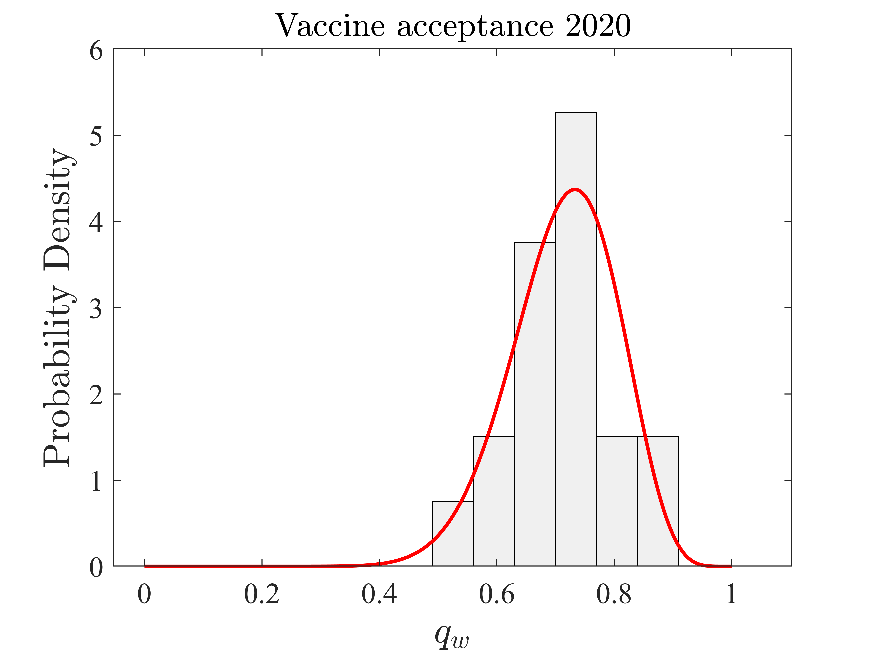}
            \caption{}
            \label{}
        \end{subfigure}
        \begin{subfigure}{.45\textwidth}
            \centering
            \includegraphics[scale=0.55]{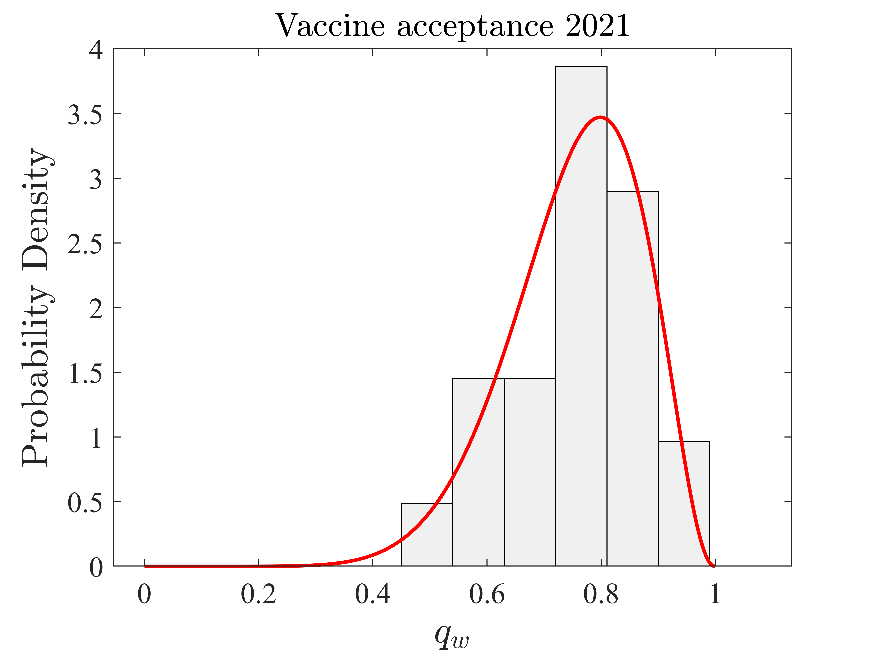}
            \caption{}
            \label{}
        \end{subfigure}
        \begin{subfigure}{.45\textwidth}
            \centering
            \includegraphics[scale=0.55]{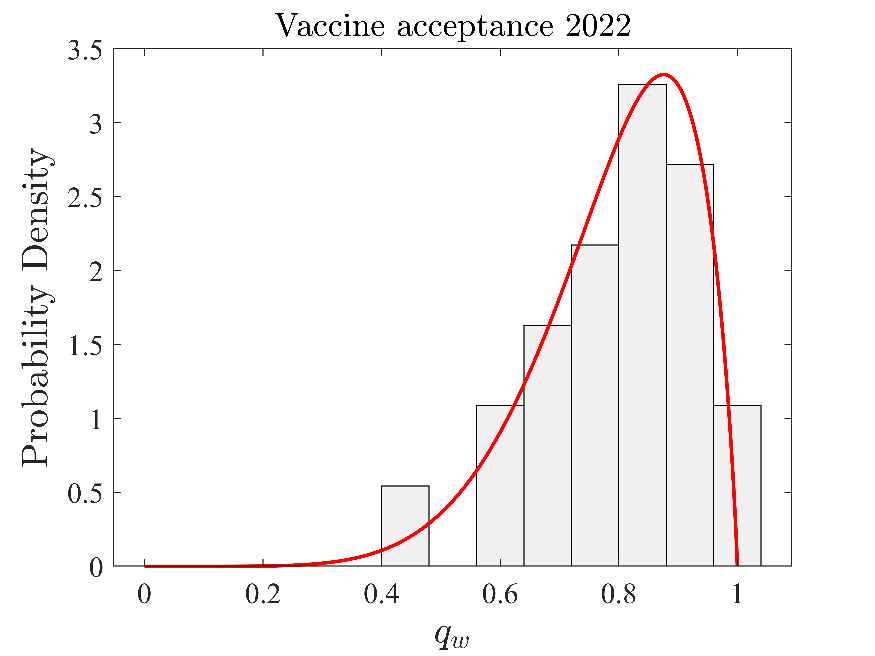}
            \caption{}
            \label{}
        \end{subfigure}
        \begin{subfigure}{.45\textwidth}
            \centering
            \includegraphics[scale=0.55]{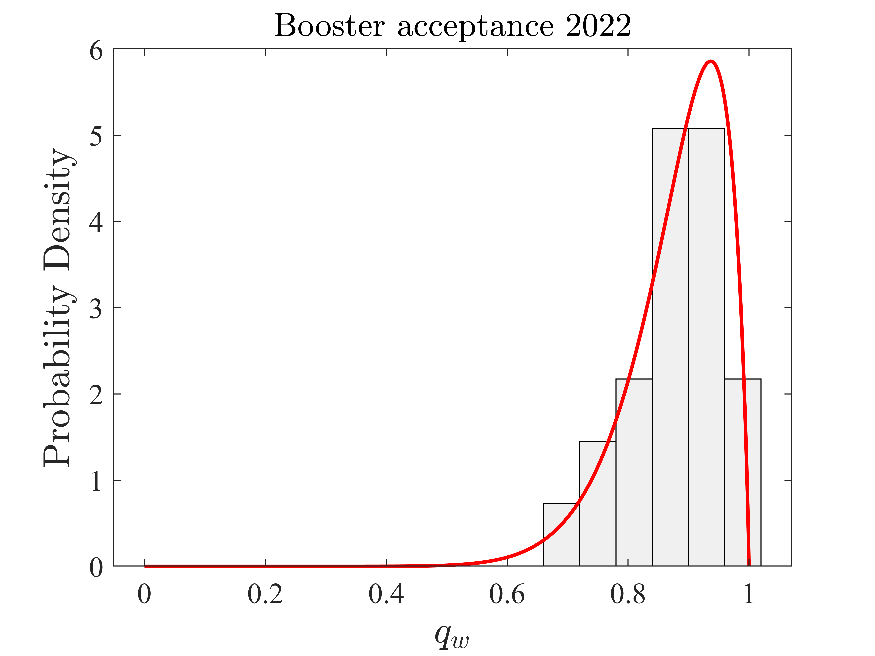}
            \caption{}
            \label{}
        \end{subfigure}
        \caption{Normalized histograms representing the probability density of vaccine willingness as a function of $q_w$, the fraction of the population with a positive opinion towards vaccination, fitted with Beta distributions (red solid curves) from the data of COVID-19 vaccine willingness in 23 countries for the years 2021 and 2022 (19 countries for the year 2020), including a vaccine booster dose in 2022. The bin sizes of the histograms are 0.08, 0.06, 0.09, and 0.07, respectively.}
        \label{fig:vac_opinion_global}
    \end{figure*}

    Lazarus \emph{et al.} \cite{lazarus2023} have carried out an extensive survey on VH among 23,000 respondents in 23 countries (1000 respondents from each country) during the COVID-19 pandemic years of 2020, 2021, and 2022. This study presents the percentagewise data on vaccine willingness among respondents from 2020 to 2022, including vaccine willingness for the booster dose in 2022. We have assumed that the data from this survey represent the situation of VH in the entire country. For our analysis, these percentagewise data are converted into fractional data ranging from 0 to 1. Next, these data are used to plot normalized histograms representing the probability density of vaccine willingness as a function of $q_w$ (the fraction of vaccine-willing individuals in the total population)  with suitable bin sizes,  for the years 2020-2022 including vaccine willingness for the 2022 booster dose (Fig.~\ref{fig:vac_opinion_global}). The histograms are fitted by the Beta distribution with the fitted parameter values given in Table~\ref{tab:beta_tab_1}. The data are also fitted by other distributions like Normal, Log-Normal, and Weibull with the ``\verb|fitdist|" function of MATLAB. This function fits the distribution to the data using the maximum likelihood estimation (MLE) technique. The goodness-of-fit for the different distributions is computed using the Kolmogorov-Smirnov test (KS test) \cite{stephens1947,berger2014,simrad2011}.
    
    In the One-sample KS test, a comparison between a sample distribution, based on collected data (COVID-19 data), and a known theoretical distribution (Normal, Weibull, Beta or Log-Normal) is carried out to obtain a numerical estimate of the goodness-of-fit of the two distributions. For this purpose, one needs to carry out the following computations. In the first step, one calculates the quantity

    \begin{equation}
        D=\mathrm{sup}_x\left|F(x)-F_{n}(x)\right|,
    \end{equation}

    \noindent where $\mathrm{sup}$ stands for the supremum, i.e., the largest value over all possible values of $x$. $F_n (x)$ is the empirical cumulative distribution function of the sample calculated as the proportion of sample data points that are less than or equal to $x$ with $n$ being the number of data points in the sample and $F(x)$ is the cumulative distribution function for the reference theoretical distribution. The $D$ statistic or the KS-test statistic provides a measure of the maximum vertical distance between the two cumulative distribution functions. A high value of $D$ indicates significant discrepancies between the two distributions. Conversely, a lower value of $D$ shows that the empirical and reference distributions are closely aligned.
    
    The next step in the KS test involves the checking of two hypotheses: the Null and the Alternative. The Null Hypothesis rests on the assumption that the empirical distribution is well-described by the theoretical distribution, whereas in the Alternative Hypothesis, the two distributions are dissimilar. To check which hypothesis is true, one can  find  a critical  $D$ value from the KS Tables or by using computational software. One can also derive the “$p$ values” from a knowledge of the $D$ values and the sample size. The Null Hypothesis is rejected if the $D$ value is greater than the critical value or the $p$ value is lower than a chosen significance level, commonly assumed to be 0.05. 

    \begin{table}
        \caption{Parameters of the Beta distributions in Fig.~\ref{fig:vac_opinion_global}.}
        \centering
        \begin{tabular}{|c|c|c|c|c|c|}
            \hline
            Type & Year  & $\epsilon_{1}$ & $\epsilon_{2}$ \\
            \hline
            COVID-19 Vaccine & 2020 & 17.25 & 6.92 \\
            COVID-19 Vaccine & 2021 & 9.85 & 3.23 \\
            COVID-19 Vaccine & 2022 & 7.12 & 1.87 \\
            COVID-19 Booster & 2022 & 13.53 & 1.85 \\
            \hline
        \end{tabular}
        
        \label{tab:beta_tab_1}
    \end{table}

    \begin{table}
        \caption{Computed KS-test goodness-of-fit values and corresponding $p$ values for the different distributions used in the fitting of the histograms plotted in Fig.~\ref{fig:vac_opinion_global}.}
        \centering
        \begin{tabular}{|c|c|c|c|c|}
            \hline
            Type & Year  & Distributions & KS-test statistic & $p$ values\\
            \hline
            \multirow{4}{*}{\makecell{COVID-19 \\ Vaccine}} & \multirow{4}{*}{2020} & Beta & 0.1029 & 0.9751 \\
            \cline{3-5}
            & & Normal & 0.0920 & 0.9923\\
            \cline{3-5}
            & & Log-Normal & 0.1115 & 0.9515\\
            \cline{3-5}
            & & Weibull & 0.1120 & 0.9496\\
            \hline
            \multirow{4}{*}{\makecell{COVID-19 \\ Vaccine}} & \multirow{4}{*}{2021} & Beta & 0.1269 & 0.8080\\
            \cline{3-5}
            & & Normal & 0.1399 & 0.7075\\
            \cline{3-5}
            & & Log-Normal & 0.1710 & 0.4614\\
            \cline{3-5}
            & & Weibull & 0.1034 & 0.9455\\
            \hline
            \multirow{4}{*}{\makecell{COVID-19 \\ Vaccine}} & \multirow{4}{*}{2022} & Beta & 0.0951 & 0.9726\\
            \cline{3-5}
            & & Normal & 0.1202 & 0.8550\\
            \cline{3-5}
            & & Log-Normal & 0.1544 & 0.5897\\
            \cline{3-5}
            & & Weibull & 0.1142 & 0.8922\\
            \hline
            \multirow{4}{*}{\makecell{COVID-19 \\ Booster}} & \multirow{4}{*}{2022} & Beta & 0.0818 & 0.9944\\
            \cline{3-5}
            & & Normal & 0.1375 & 0.7266\\
            \cline{3-5}
            & & Log-Normal & 0.1564 & 0.5736\\
            \cline{3-5}
            & & Weibull & 0.0858 & 0.9903\\
            \hline
        \end{tabular}
        
        \label{tab:All_dist_fit_opinion}
    \end{table}

    The results of the  KS test, obtained for the different distributions, are shown in Table~\ref{tab:All_dist_fit_opinion}. The ``\verb|kstest|" function of MATLAB has been used to compute the KS-test statistic and the associated $p$ value for each reference distribution. The $p$ values of all the fits are much above the significance level 0.05 and thus we cannot reject the Null Hypothesis, implying that all of these distributions can be used and fit the data well. Also, a lower computed value of the KS-test statistic indicates a better fit among all the fitted distributions. The best fitting distributions in the years 2020 and 2021 are the Normal and Weibull distributions, respectively. The data for the year 2022 (both vaccine and booster) are best fitted by the Beta distribution.

\begin{figure*}
    \centering
        \begin{subfigure}{.45\textwidth}
            \centering
            \includegraphics[scale=0.55]{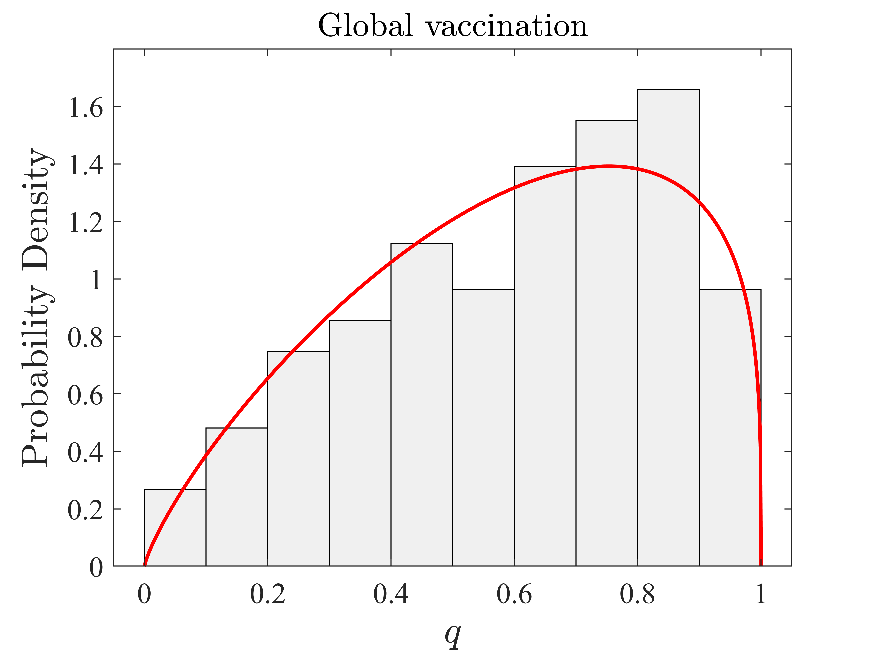}
            \caption{}
            \label{fig:global_vac_all}
        \end{subfigure}
        \begin{subfigure}{.45\textwidth}
            \centering
            \includegraphics[scale=0.55]{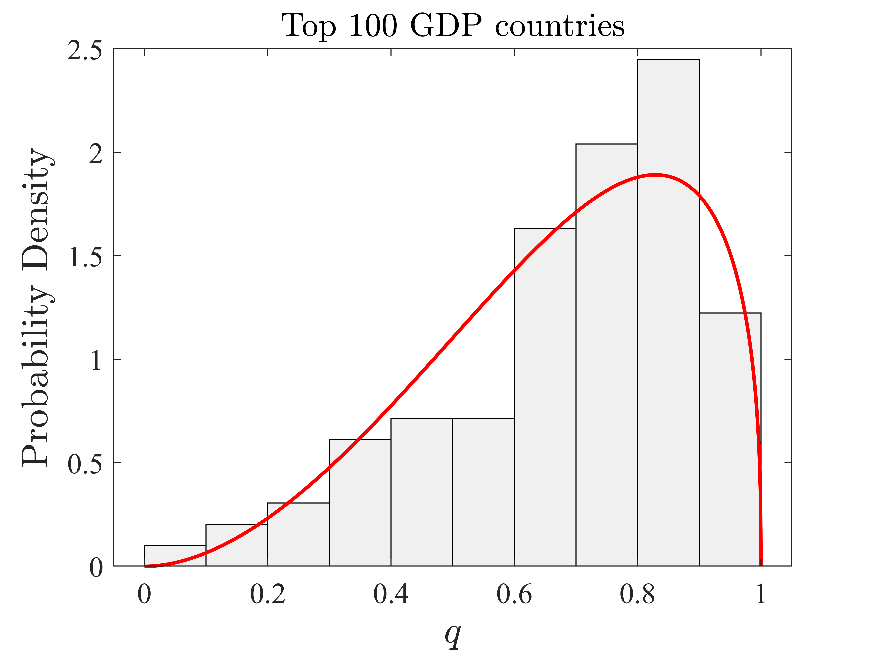}
            \caption{}
            \label{fig:global_vac_100gdp}
        \end{subfigure}
        \caption{(a) Distribution of the proportion of the population with at least one dose of the COVID-19 vaccine globally (b) the same distribution for the top 100 GDP countries.}
        \label{fig:global_vac}
    \end{figure*}

     VH is a complex phenomenon with previous studies of influenza VH reporting more than 70 factors which influence VH \cite{lazarus2023}. The Beta distribution, as obtained from the VH model, is an outcome of coexisting individual and group propensities for vaccine rejection/acceptance. From Table~\ref{tab:beta_tab_1}, the fitted values for the parameters, $\epsilon_{1}=\mfrac{\sigma_{1}}{h}$, $\epsilon_{2}=\mfrac{\sigma_{2}}{h}$, for vaccine acceptance and rejection, respectively, indicate that individual propensities are dominant over the herding mechanism (both $\epsilon_{1},\epsilon_{2}>1$) and also  $\epsilon_{1}>\epsilon_{2}$. The values of both the parameters decrease in the succeeding years indicating a greater contribution from the herding mechanism. Yang \emph{et al.} \cite{yang2024} have analyzed vaccination data from the global COVID-19 pandemic and shown that the Axios-Ipsos survey, based on the beta regression model, exhibited a strong correlation between VH and actual vaccine uptake. Based on this evidence, we assume that the fraction of the vaccinated population in the total population is also given by the Beta distribution. To analyze this assumption, we considered the countrywise vaccination progress data from the Johns Hopkins University \& Medicine Coronavirus Resource Center \cite{JHU_website}, which provides the percentage of the population that had received at least one dose of the vaccine. The adjusted percentagewise dataset is then converted into the fractional dataset and the histogram is plotted in Fig.~\sref{fig:global_vac}{fig:global_vac_all} along with the fitted Beta distribution, where $q$ represents the fraction of the population who had received at least one dose of the vaccine. Taking into account the fact that economically weak countries cannot afford to undertake a full-fledged vaccination campaign, Fig.~\sref{fig:global_vac}{fig:global_vac_100gdp} shows the vaccine-accepted fraction of the population, from the top 100 GDP (Gross Domestic Product) countries, versus $q$, fitted also by the Beta distribution.

    Table~\ref{tab:All_dist_fit_global} shows the KS-test statistic and $p$ values for the different distribution fits of the global vaccination data. The Beta distribution does not reject the null hypothesis for both the global vaccination data and the top 100 GDP countries data (as $p\gg 0.05$), whereas the other distributions reject the null hypothesis or barely accept it for at least one of these datasets.  In addition, the Beta distribution has the lowest value of KS-test statistic for both datasets. Based on this evidence, we conclude that both datasets are best represented by the Beta distribution.

    \begin{table}
        \caption{KS-test statistic values and $p$ values for the different distributions fit on the global vaccination data (Fig.~\ref{fig:global_vac}).}
        \centering
        \begin{tabular}{|c|c|c|c|}
            \hline
            Type  & Distributions & KS-test statistic & $p$ values \\
            \hline
            \multirow{4}{*}{\makecell{COVID-19 \\ Vaccine data \\ of all countries}} & Beta & 0.0518 & 0.6767\\
            \cline{2-4}
            & Normal & 0.0900 & 0.0910\\
            \cline{2-4}
            & Log-Normal & 0.1796 & $9.4976\times 10^{-6}$\\
            \cline{2-4}
            & Weibull & 0.1014 & 0.0399\\
            \hline
            \multirow{4}{*}{\makecell{COVID-19 \\ Vaccine data \\ for top 100 \\ GDP countries}} & Beta & 0.0874 & 0.4579\\
            \cline{2-4}
            & Normal & 0.1156 & 0.1348\\
            \cline{2-4}
            & Log-Normal & 0.1973 & $8.0795\times 10^{-4}$\\
            \cline{2-4}
            & Weibull & 0.1147 & 0.1401\\
            \hline
        \end{tabular}
        
        \label{tab:All_dist_fit_global}
    \end{table}

    The choice of the Beta distribution in fitting COVID-19 vaccine data is motivated by the fact that the distribution is a popular choice in the modeling of data based on Bayesian statistics \cite{van2021,eddy2004,gao2020}. The procedure is based on the Bayes’ rule

    \begin{equation}
        p(\theta|y)=\frac{p(y|\theta)p(\theta)}{p(y)},
    \end{equation}
    
    \noindent where $p(\theta|y)$ and $p(\theta)$ are the posterior and prior distributions of an unknown parameter $\theta$. For example, $\theta$ could be the probability that an individual in a population catches the COVID-19 infection. The likelihood function $p(y|\theta)$ describes the conditional probability of the observed data $y$ given the model parameter $\theta$. For example, if from actual observation, one gets to know that $k$ persons out of $n$ individuals fall sick, the likelihood of the event is given by

    \begin{equation}
        p(y|\theta)\sim \binom nk \theta^k (1-\theta)^{n-k},
    \end{equation}

    \noindent which is the well-known Binomial distribution. The prior distribution is generally chosen before data collection. The posterior distribution, a conditional distribution, provides an updated knowledge of the unknown parameter $\theta$, on combining the prior distribution with the likelihood function based on collected data. The Beta distribution, as a prior distribution, has a special relationship with the Binomial likelihood function. The combination of the two yields a posterior distribution which is another Beta distribution, with renormalized exponents $\epsilon_{1}$ and $\epsilon_{2}$ in Eq.~\pref{eq:kirman_pst}. This unique feature, along with the flexibility of the Beta distribution, make the distribution a convenient choice for the prior distribution. The Bayesian analysis has been applied successfully across diverse disciplines \cite{van2021,eddy2004,gao2020}, including medicine and vaccine development.

\section{Epidemic model with vaccination}\label{Sec:epi_model_vac}

    In the previous section, based on COVID-19 vaccination data, the Beta distribution is found to provide a good description of the distribution of the vaccinated fraction $q$ of a population. In a deterministic description, vaccination is taken into account,  in the SIS model, by rewriting the rate equation (Eq.~\pref{eq:y_main}) as \cite{oregan2013}

    \begin{equation} \label{i_eq_vac}
        \frac{dy}{dt}=f(y)=\beta (1-q) y(1-y)+\eta (1-y) -\gamma y.
    \end{equation}

    To understand the effect of vaccination, we consider the simpler case of zero immigration rate, i.e., $\eta=0$. The fixed points of the dynamics are then  $y^*= 0$ and  $y^*=1-\mfrac{1}{R},~R\geq 1$ where $R=R_{0}(1-q)$ is the effective reproduction number and $R_{0}=\mfrac{\beta}{\gamma}$ is the basic reproduction number. The system dynamics now have a transcritical bifurcation at $R=1$ that is reached for larger values of $R_{0}=\mfrac{1}{1-q}>1$ as $q$ increases indicating the favorable effect of vaccination on the total eradication of the infection.

    We now introduce uncertainty in the deterministic dynamics by assuming the fraction of vaccinated population $q$ to be beta distributed, the justification for which has been provided in the preceding section, i.e., the steady-state PDF for $q$ has the form given in Eq.~\pref{eq:kirman_pst} with $x$ replaced by $q$. The probability distribution of $R=R_0(1-q)$ is then expressed as 


    \begin{multline}\label{eq:pdf_R}
        p_{\text{st}}(R)=\frac{1}{R_{0}B(\epsilon_{1},\epsilon_{2})}\left(1-\frac{R}{R_{0}}\right)^{\epsilon_{1}-1}\left(\frac{R}{R_{0}}\right)^{\epsilon_{2}-1}, \\ 0\leq R\leq R_{0}.
    \end{multline}

    \noindent where $B(\epsilon_{1},\epsilon_{2})=\mfrac{\Gamma(\epsilon_{1})\Gamma(\epsilon_{2})}{\Gamma(\epsilon_{1}+\epsilon_{2})}$. A similar probability distribution, for the basic reproduction number  $R_0$, appears in a Bayesian inferential approach to study the progress of the H1N1 influenza epidemic \cite{kadi2015}. In the case of the SIS model with Beta-distributed vaccination, the endemic equilibrium $y^*=1-\mfrac{1}{R}$ exists in the range $1\leq R\leq R_{0}$. Thus, $\int_{1}^{R_0}p_{\text{st}}(R)dR$ provides a measure of the probability that the population is not completely free from the disease. Similarly, $\int_{0}^{1}p_{\text{st}}(R)dR$ is the probability of eradication of the disease from the population. Defining $r=R/R_{0}$, the probability distribution of $r$ can be written as

    \begin{equation}\label{pdf_r}
        p_{\text{st}}(r)=\frac{1}{B(\epsilon_{1},\epsilon_{2})}\left(1-r\right)^{\epsilon_{1}-1}r^{\epsilon_{2}-1}, \qquad 0\leq r\leq 1.
    \end{equation}

\begin{figure*}
	\centering
	  \begin{subfigure}{.45\textwidth}
	      \centering
	      \includegraphics[scale=0.5]{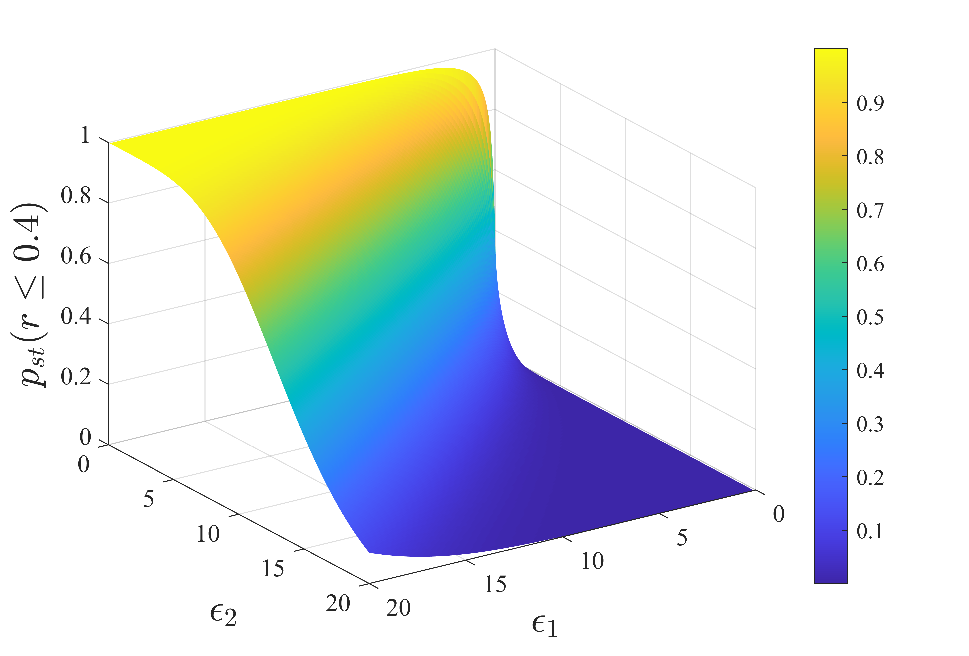}
	      \caption{}
	      \label{fig:3D_R_25}
	  \end{subfigure}
	  \begin{subfigure}{.45\textwidth}
	      \centering
	      \includegraphics[scale=0.5]{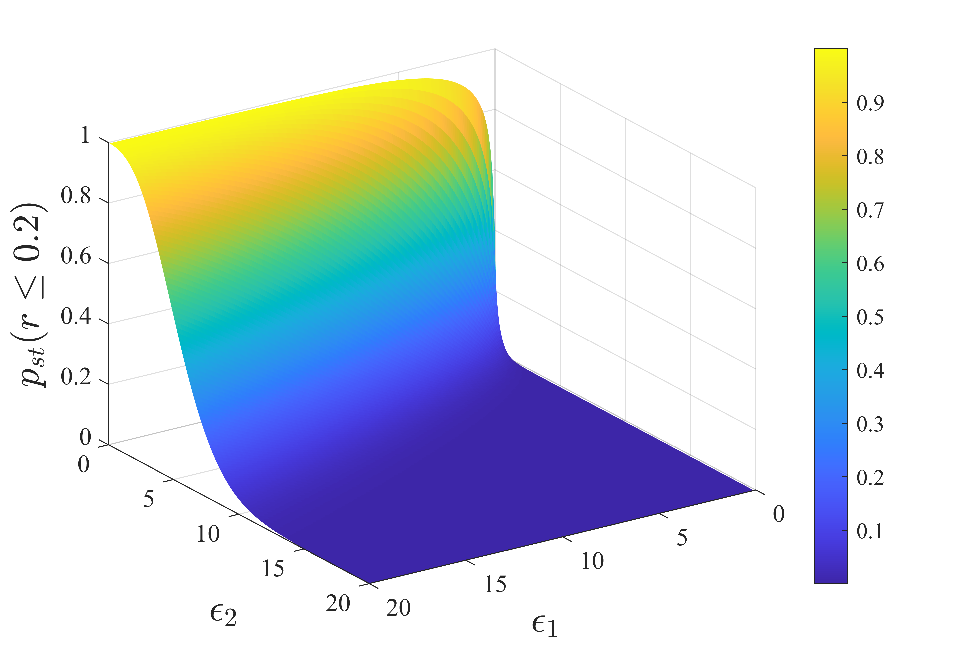}
	      \caption{}
	      \label{fig:3D_R_5}
	  \end{subfigure}
	  \caption{Surface plot of $p_{\text{st}}(r\leq r_{\text{th}})$ (Eq.~\pref{eq:pst_r_leq_rth} as a function of the parameters $\epsilon_1$ and $\epsilon_2$) for (a) $r_{\text{th}}=0.4$ and (b) $r_{\text{th}}=0.2$.}
	  \label{fig:3D_plots}
	\end{figure*}

    Let the threshold value of $r$ for disease-free condition be denoted as $r_{\text{th}}$. Then

    \begin{multline}\label{eq:pst_r_leq_rth}
        p_{\text{st}}(r\leq r_{\text{th}})=\frac{1}{B(\epsilon_{1},\epsilon_{2})}\int_{0}^{r_{\text{th}}}\left(1-r\right)^{\epsilon_{1}-1}r^{\epsilon_{2}-1}dr \\ =\frac{B(r_{\text{th}};\epsilon_{2},\epsilon_{1})}{B(\epsilon_{1},\epsilon_{2})},
    \end{multline}

    where $B(r_{\text{th}};\epsilon_{2},\epsilon_{1})=\mfrac{1}{\epsilon_{2}}r_{\text{th}}^{\epsilon_{2}}\left\{_{2}F_{1}\left(1-\epsilon_{1},\epsilon_{2},1+\epsilon_{2};r_{\text{th}}\right)\right\}$ with $\mfrac{1}{\epsilon_{2}}r_{\text{th}}^{\epsilon_{2}}\left\{_{2}F_{1}\left(1-\epsilon_{1},\epsilon_{2},1+\epsilon_{2};r_{\text{th}}\right)\right\}$ being the hypergeometric function. Figure~\ref{fig:3D_plots} shows the three-dimensional plot of  $p_{\text{st}}$ $(r\leq r_{\text{th}} )$ versus the parameters $\epsilon_{1}$ and $\epsilon_{2}$ for  $R_0=$ 2.5 and 5.0, i.e., $r_{\text{th}}=$ 0.4 and 0.2, respectively. The plot shows that the probability for the disease-free condition $(p_{\text{st}} (r\leq r_{\text{th}}))$ is higher when the value of $\epsilon_{1}$  is substantially larger than the value of $\epsilon_{2}$  (more vaccine acceptance). The probability $p_{\text{st}}\rightarrow 1$ as $\epsilon_{2}\rightarrow  0$ for a range of values of $\epsilon_{1}$. As the value of $\epsilon_{2}$ increases, the probability of achieving the disease-free condition $(p_{\text{st}} (r\leq r_{\text{th}}))$ decreases rapidly as more people are hesitant to take the vaccine. Therefore, a higher value of $\epsilon_{1}$ relative to $\epsilon_{2}$ is a prerequisite to achieve a disease-free condition. 

    With the inclusion of vaccination in the stochastic SIS model (Sec.~\ref{Sec:sto_w_vac}), the parameters $R_0$ and $V$ are renormalized as $R=R_0 (1-q)$ and $V_R=\mfrac{V}{1-q}$ where $q$ is the fraction of vaccinated population. The bifurcation diagrams of the stochastic model in terms of the new parameters $R$ and $V_R$ have the same appearance as in the absence of vaccination. We now assume that the parameter $q$ has a distribution, namely, the Beta distribution, instead of a fixed value. This results in a distribution of the effective reproduction number $R$ (Eq.~\pref{eq:pdf_R}). In the usual bifurcation theory, the bifurcation parameter has a known single value. When the parameter has a probability distribution, the bifurcation diagram loses its simple representation with the boundary lines separating different dynamical regimes becoming probabilistic since each value of the parameter can give rise to different outcomes with specific probabilities. In summary, when the vaccination parameter $q$ is deterministic, one can identify distinct dynamical regimes characterized by noise-induced transitions and describe critical behavior in terms of universality classes. When $q$ has a probability distribution, the outcomes for a single value are also probabilistic.

\section{Concluding Remarks}\label{Sec:conclusion}

    The study of epidemic and vaccine models has acquired special importance in recent years due to the widespread occurrence of infectious diseases such as H1N1 influenza and COVID-19. The major issues in focus are the dynamics of epidemic spread and the adoption of control strategies such as vaccination, social distancing and intermittent periods of lock-down. The SIS model is the simplest model of epidemic spread and its utility lies in highlighting at the most basic level the key features of epidemic dynamics, namely, the transmission of infection through contacts and the existence of an absorbing phase transition, occurring at a critical value of the basic reproduction number $R_0$, from a phase of active infection to an absorbing phase in which the infection is totally eradicated. While diseases such as tuberculosis, meningitis, and gonorrhea \cite{mamis2023,keeling2008,brauer2012,allen2008} are known to follow the SIS pattern of dynamics (no permanent immunity), the SIS model is of broader relevance due to its close links with other nonequilibrium statistical models exhibiting absorbing phase transitions. Our study focuses on the stochastic SIS model that includes immigration. We have shown that the model exhibits a critical-point transition from a region of bimodality to that of unimodality with the critical transition belonging to the universality class of the mean-field Ising model. The SIS model with immigration shares universal features with the HL and other related models \cite{horsthemke1984,arnold1978,bose2022,erez2019,bose2019,bose2022criticality}, such as noise-induced first-order transitions accompanied by hysteresis and critical point transitions belonging to the Ising mean-field universality class. 

    Any vaccination program aimed at the eradication of infection is beset by the problem of VH among a sizable fraction of the population. Our model of VH is developed along the lines of Kirman’s ant model \cite{kirman1993} in which the decision-making process involves a random binary choice, either decide individually or be guided by the herding instinct. The steady-state distribution of the vaccine-willing (or vaccine-hesitant) fraction of the population is given by the Beta distribution which turns out to be the universal distribution for a number of statistical physics problems involving a random binary choice \cite{kirman1993,alfarano2005,kononovicius2018}. The Beta distribution is shown to give quite a good fit to the COVID-19 data on VH \cite{lazarus2023} (Fig.~\ref{fig:vac_opinion_global}). Based on a survey of COVID-19 vaccination data \cite{yang2024}, the Beta distribution is found to give a good fit also to the global data for the fraction of vaccinated population (Fig.~\ref{fig:global_vac}). In most studies on vaccine acceptance and efficacy, the standard statistical procedure adopted is that of Bayesian analysis \cite{wang2021} in which a likelihood function improves upon an assumed prior distribution to yield a final posterior distribution. In many such studies \cite{yang2024}, the Beta distribution is a popular choice for the initial prior distribution. Our VH model illustrates a possible physical origin of the Beta distribution based on random binary choice.  

    The conceptual basis of epidemiological studies has now been extended to the study of how viruses spread in a cell population within a host \cite{williams2024,pearson2011}. The stochastic viral dynamics have several interesting features adding new aspects to the spreading process. The basic reproduction number $R_0$ is a random variable and defined to be “the number of new cells infected by one initial infected cell in an otherwise susceptible (target cell) population”. The variability in $R_0$ is partly associated with the number of viral progeny generated by an infected cell during its lifetime, i.e., the randomness has a cellular origin. A virus entering a host cell captures the cellular machinery to multiply resulting eventually in the viruses exiting the host cell in a burst and killing the cell in the process or exiting the cell continuously through a process of “budding” during the lifetime of the infected cell. While the mode of viral production makes the task of model building more complex, the stochastic viral dynamics shares an important feature with the stochastic SIS model. In both cases, the deterministic model predicts that an infection is surely established when $R_0>1$. In the stochastic versions of both models, there is a finite probability that the infection is totally eradicated even if $R_0>1$. The probability depends both on the mean value of  $R_0$ as well as its probability distribution (Fig.~\ref{fig:3D_plots}). Williams \emph{et al.} \cite{williams2024} point out that by focusing exclusively on the mean value of   $R_0$, one may fail to capture the important aspects of the early infection dynamics. In summary, our study highlights the universal features of epidemic and vaccine models, shared with other nonequilibrium statistical physics models identifying thereby a common framework of conceptual understanding as well as calculational schemes.
\vspace{12pt}
\section*{Acknowledgement}

S. C. thanks the University Grant Commission (UGC), Government of India for providing financial support through ``CSIR-UGC NET-JRF". S. R. acknowledges the financial assistance provided by St. Xavier's College (Autonomous), Kolkata in the form of `Intramural Research Grant for R\&D Projects' (Sanction No. IMSXC2022-23/005). I. B. acknowledges the support of NASI, Allahabad, India, under the Honorary Scientist Scheme.

\section*{Author Contributions}

S. C. has done the calculations and written the program codes for this work. I. B. helped in conceptualization/physical understanding and implementation of calculational schemes. S. R., along with S. C. has helped in conceptualizing portions of the manuscript and also planned and executed schemes for different calculations and plots.  All of the authors contributed to the development of this work and the preparation of this manuscript. Also, all of the authors have thoroughly checked and reviewed this manuscript before submitting it.  

\bibliography{References_aps_w}
    
\end{document}